\documentclass{article}
\pdfoutput=1
\usepackage{arxiv}

\usepackage[utf8]{inputenc} 
\usepackage[T1]{fontenc}   
\usepackage{hyperref}       
\usepackage{url}            
\usepackage{booktabs}      
\usepackage{amsfonts}       
\usepackage{nicefrac}      
\usepackage{microtype}

\usepackage[ruled]{algorithm2e} 
\usepackage{epsfig,subfigure}
\usepackage[noend]{algorithmic}
\usepackage{amsmath,amsthm}
\usepackage{amssymb}
\usepackage{enumitem}

\newtheorem{definition}{Definition}
\newtheorem{example}{Example}
\newtheorem{problem}{Problem}
\newtheorem{lemma}{Lemma}

\newcommand{\bas}{\mathit{DTJb}}
\newcommand{\rep}{\mathit{DTJr}}
\newcommand{\ind}{\mathit{DTJi}}
\newcommand{\prob}{{Distributed Subtrajectory Join}}

\usepackage{setspace}
\title{Distributed Subtrajectory Join on Massive Datasets}

\author{
  Panagiotis Tampakis$^1$, Christos Doulkeridis$^2$, Nikos Pelekis$^3$ and Yannis Theodoridis$^1$\\
  $^1$Department of Informatics\\
  $^2$Department of of Digital Systems\\
  $^3$Department of Statistics \&\ Insurance Science\\
  University of Piraeus\\
  Piraeus, Greece \\
  \texttt{\{ptampak,cdoulk,npelekis,ytheod\}@unipi.gr} \\
}

\begin{document}
\maketitle

\begin{abstract}
Joining trajectory datasets is a significant operation in mobility data analytics and the cornerstone of various methods that aim to extract knowledge out of them.
In the era of Big Data, the production of mobility data has become massive and, consequently, performing such an operation in a centralized way is not feasible. 
In this paper, we address the problem of \emph{\prob} processing by utilizing the MapReduce programming model. 
Compared to traditional trajectory join queries, this problem is even more challenging since the goal is to retrieve all the ``maximal'' portions of trajectories that are ``similar''.
We propose three solutions: (i) a well-designed basic solution, coined $\bas$, (ii) a solution that uses a preprocessing step that repartitions the data, labeled $\rep$, and (iii) a solution that, additionally, employs an indexing scheme, named $\ind$. In our experimental study, we utilize a 56GB dataset of real trajectories from the maritime domain, which, to the best of our knowledge, is the largest real dataset used for experimentation in the literature of trajectory data management. 
The results show that $\ind$ performs up to 16$\times$ faster compared with $\bas$, 10$\times$ faster than $\rep$ and 3$\times$ faster than the closest related state of the art algorithm.  
\end{abstract}


\maketitle

\section{Introduction} \label{sec_intro}

During the recent years, the proliferation of GPS enabled devices has led to the production of enormous amounts of mobility data. 
This ``explosion'' of data generation has posed new challenges in the world of mobility data management. One of these challenges is the so-called trajectory join problem, which aims to find all pairs of ``similar'' (i.e. nearby in space-time) trajectories in a dataset~\cite{DBLP_conf/gsn/BakalovT06,DBLP_conf/gis/ChenP09,DBLP_conf/time/DingTS08,DBLP_journals/tkde/TaLXLHF17}. An even more interesting and challenging problem is the subtrajectory join query~\cite{DBLP_conf/mdm/BakalovHKT05}, where, for each pair of trajectories, we want to retrieve all the ``portions'' of trajectories that are ``similar''. However, the subtrajectory join is a processing-intensive operation. 
Centralized algorithms do not scale with the size of today's trajectory data, thus parallel and distributed algorithms are necessary in order to provide efficient processing of subtrajectory join, an issue largely overlooked in the related research.

Several modern applications that manage trajectory data could benefit from such an operation. For instance, in the urban traffic domain, carpooling is becoming increasingly popular. More concretely, consider a mobile application which tries to match users that can share a ride based on their past movements. Here, given a set of trajectories we want to find all the pairs of users that can share a ride for a portion of their everyday routes without significantly deviating (spatially and temporally) from their daily routine (i.e. retrieve all pairs of maximal subtrajectories that move close in space and time). Another interesting scenario concerns the identification of suspicious movement by a governmental security agency. For instance, given a set of trajectories that depict the movement of suspicious individuals, we would like to retrieve all the pairs of moving objects that move ``close'' to each other for more than a threshold (moving together for small periods of time could be considered as coincidental) as candidates for illegal activity.
Moreover, such a query is in fact the building block for a number of operations than aim to identify mobility patterns, such as co-movement patterns (e.g. flocks~\cite{DBLP_conf/gis/GudmundssonK06}, convoys~\cite{DBLP_journals/pvldb/JeungYZJS08}, swarms~\cite{DBLP_journals/pvldb/LiDHK10}). 
An even more challenging problem is that of subtrajectory clustering ~\cite{DBLP_conf/edbt/PelekisTVPT17,DBLP_conf/pods/AgarwalFMNPT18}. An interesting application scenario of subtrajectory clustering is network discovery, where given a set of trajectories (e.g from the maritime or the aviation domain) we want to identify the underlying network of movement by grouping subtrajectories that move ``close'' to each other and use cluster representatives/medoids as network edges. One of the main goals of subtrajectory clustering is to segment trajectories to subtrajectories. Finally, trajectory segmentation techniques~\cite{DBLP_journals/tkde/PanagiotakisPKRT12,DBLP_conf/edbt/PelekisTVPT17}, can directly benefit from the subtrajectory join query since their input, for each trajectory, is the number of objects that were located close to it at any given time.
However, the bottleneck in all these applications is the underlying processing cost of the join operation, which calls for parallel and distributed solutions that scale beyond the limitations of a single machine.

\begin{figure} [thb]
    \begin{minipage}[t]{.5\linewidth}
      \centering
          \includegraphics[width=1\textwidth]{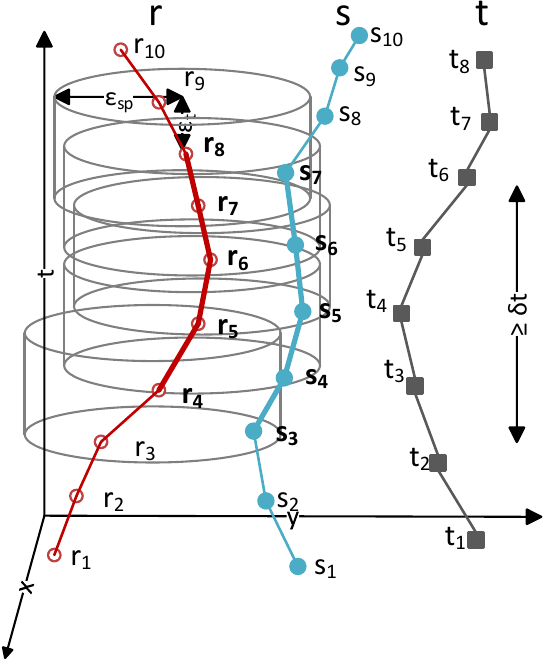}
    \end{minipage}%
    \begin{minipage}[t]{.5\linewidth}
      \centering
          \includegraphics[width=1\textwidth]{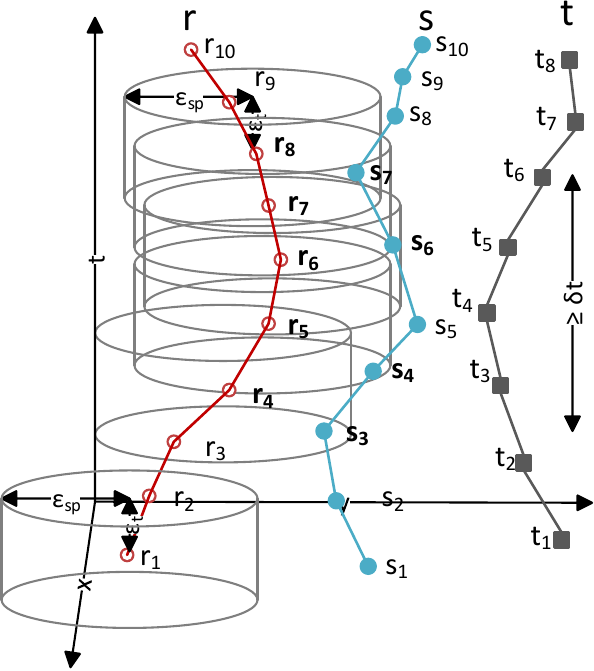}
    \end{minipage}
  \caption{(a) A pair of maximally ``matching'' subtrajectories and (b) a breaking point $r_{1}$ and a non-joining point $s_{5}$ w.r.t. $r$.}
  \label{fig_intro}
\end{figure}

Inspired by the above application scenarios, the problem that we address in this paper is as follows: given two sets of trajectories (or a single set and its mirror in the case of self-join), identify all pairs of maximal ``portions'' of trajectories (or else, subtrajectories) that move close in time and space w.r.t a spatial threshold $\epsilon_{sp}$ and a temporal tolerance $\epsilon_t$, for at least some time duration $\delta t$. To illustrate this informal definition, as depicted in Figure~\ref{fig_intro}(a), given two trajectories $r$ and $s$, the pair of their maximal matching ``portions'' is ($\{r_4, r_5, r_6, r_7, r_8\}, \{s_3, s_4, s_5, s_6, s_7\}$). Each point of a trajectory defines a spatio-temporal 'neighborhood' area around it, a cylinder of radius $\epsilon_{sp}$ and height $\epsilon_t$. In order for a pair of subtrajectories to be considered ``matching'', each point of a subtrajectory must have at least one point of the other subtrajectory in its ``neighborhood'', thus making the result symmetrical. A pair of matching subtrajectories is maximal if there exists no superset of either subtrajectories that can replace them and the pair still qualifies as a ``matching'' pair.

There have been some efforts to tackle variations of this problem in a centralized way~\cite{DBLP_conf/mdm/BakalovHKT05,DBLP_conf/gsn/BakalovT06,DBLP_conf/gis/ChenP09}. However, these solutions discover pairs of entire trajectories and cannot identify matching sub-trajectories. In~\cite{DBLP_conf/gis/BakalovHT05}, all pairs of ``matching'' (w.r.t a spatial threshold) subtrajectories of exactly $\delta t$ duration are retrieved in contrast with the problem addressed in this paper, where the goal is to identify maximally ``matching'' subtrajectories, which is vital for exploiting the output in subsequent steps, e.g. the mining operations mentioned above. Moreover, applying these centralized solutions to a parallel and distributed environment is not straightforward and is often impossible if radical changes to the methods/algorithms do not take place, since there are several non-trivial issues that arise. For instance, how to partition the data in such a way so that each partition can be processed independently and be of even size. 

In a recent effort, in ~\cite{DBLP_journals/vldb/ShangCWJZK18} the authors try to tackle the problem of trajectory similarity join in spatial networks in parallel. The solution proposed in ~\cite{DBLP_journals/vldb/ShangCWJZK18} handles each trajectory separately and all the data have to be replicated for each trajectory and, consequently, to each node. Due to this fact, such a solution cannot scale to terabytes of data, thus making it inapplicable to Big Data. Furthermore, such an approach assumes that the underlying network is known in advance, hence it cannot support datasets of moving objects that move freely in space (e.g. from the maritime or the aviation domain). As a result, a scenario where the goal is to identify the underlying network cannot be supported. Finally, the output of ~\cite{DBLP_journals/vldb/ShangCWJZK18} is pairs of trajectories and not subtrajectories, which is significantly different than the problem addressed in this paper. More recently, in ~\cite{DBLP_conf/sigmod/Shang0B18} the authors try to tackle the problem of trajectory similarity join. Specifically, given two sets of trajectories, a similarity function (e.g. DTW) and a similarity threshold, they aim to identify all pairs of trajectories that exceed this similarity threshold. Again, the problem addressed in~\cite{DBLP_conf/sigmod/Shang0B18} is to retrieve pairs of trajectories in contrast with the problem that we try to tackle in this paper, which is to retrieve all pairs of ``maximally matching'' subtrajectories.

It is straightforward to claim that an integral part of any algorithm that tries to address the subtrajectory join query is to identify all pairs of  points that move ``close enough'' in time and space w.r.t a spatial threshold $\epsilon_{sp}$ and a temporal tolerance $\epsilon_t$, e.g. $r_4$ and $s_3$ in Figure~\ref{fig_intro}(a). In that sense, another line of research that is closely related to our problem is that of MapReduce-based spatial~\cite{DBLP_conf/cluster/ZhangHLWX09, DBLP_journals/pvldb/AjiWVLL0S13,DBLP_conf/icde/EldawyM15} and multidimensional joins~\cite{DBLP_conf/sigmod/SilvaR12,DBLP_conf/mdm/LuoTMN12,DBLP_conf/icde/FriesBSS14}, where the goal is to identify such points. A generic solution which could form the basis for any MapReduce-based spatial (or spatiotemporal) join algorithm is presented in~\cite{DBLP_conf/cluster/ZhangHLWX09}, where the input data are partitioned into small, disjoint tiles at \emph{Map} stage and get joined at the \emph{Reduce} stage by performing a plane sweep algorithm along with a duplication avoidance technique. However, all of the above approaches try to solve a problem that is significantly different from ours since our problem is not to join spatial or multidimensional objects but identify all pairs of ``maximally matching'' subtrajectories.

In this paper, we provide efficient solutions for the \emph{\prob} processing problem, as it is formally defined in Section~\ref{sec_probl}. To the best of our knowledge, this problem has not been addressed in the literature yet.
Our main contributions are the following: 
\begin{itemize}[topsep=2pt,leftmargin= .2in]
\itemsep0em 
 \item We formally define the problem of \emph{\prob} processing, investigate its main properties, and discuss its main challenges. 
 \item We present a well-designed algorithm, called $\bas$, that solves the problem of \emph{\prob} processing by employing two MapReduce phases. 
 \item We propose an improvement of $\bas$, termed $\rep$, which is equipped with a repartitioning mechanism that achieves load balancing and collocation of temporally adjacent data. 
 \item To boost the performance of query processing even further, we introduce $\ind$, which extends $\rep$ by exploiting an indexing scheme that speeds up the computation of the join.
 \item We compare with an appropriately modified state of the art MapReduce spatial join algorithm and show that our solution performs several times better.
 \item We study the performance of the proposed algorithms by using, to the best of our knowledge, the largest real trajectory dataset (56GB) used before in the relevant literature, thus demonstrating the scalability of our algorithms. 
\end{itemize}

The rest of the paper is organized as follows. In Section~\ref{sec_probl} we introduce the problem. In Section~\ref{sec_solut}, we present $\bas$. Subsequently, in Section~\ref{sec_algor} we propose $\rep$ that utilizes a preprocessing step. In Section~\ref{sec_index}, we introduce $\ind$ that boosts the performance of the join processing. In Section~\ref{sec_exper}, we provide our experimental study. Finally, we present an overview of the related work in Section~\ref{sec_relat}, and we conclude the paper in Section~\ref{sec_concl}.

\section{Problem Statement} \label{sec_probl}

Given a set $R$ of trajectories moving in the xy-plane, a trajectory $r \in R$ is a sequence of timestamped locations $\{r_1,\dots,r_N\}$. Each $r_i = (x_i, y_i, t_i)$ represents the $i$-th sampled point, $i \in {1,\dots,N}$ of trajectory $r$, where $N$ denotes the length of $r$ (i.e. the number of points it consists of). The pair $(x_i, y_i)$ and $t_i$ denote the 2D location in the xy-plane and the time coordinate of point $r_i$ respectively. A subtrajectory $r_{i,j}$ is a subsequence $\{r_i,\dots,r_j\}$ of $r$ which represents the movement of the object between $t_i$ and $t_j$ where $i<j$.

Given a pair $(r, s)$ of trajectories (the same holds for subtrajectories) with $r \in R$ and $s \in S$, the \emph{common lifespan} $w_{r,s}$ is defined as the time interval $[max(r_1.t, s_1.t), min(r_N.t, s_M.t)]$, where $r_1$ ($s_1$) is the first sample of $r$ ($s$, respectively) and $r_N$  ($s_M$) is the last sample of $r$ ($s$, respectively). The duration of the common lifespan $w_{r,s}$ is $\Delta w_{r,s}$ = $min(r_N.t, s_M.t)$ - $max(r_1.t, s_1.t)$

Further, let $DistS(r_i, s_j)$ denote the spatial distance between two points $r_i$, $s_j$, which is defined as the Euclidean distance in this paper, even though other distance functions are also applicable. Also, let $DistT(r_i, s_j)$ denote the temporal distance, defined as $|r_i.t - s_j.t|$.

\begin{definition} \label{dfn_3}
\textbf{(Matching subtrajectories)} Given a spatial threshold $\epsilon_{sp}$, a temporal tolerance $\epsilon_t$ and a time duration $\delta t$, a ``match'' between a pair of subtrajectories $(r', s')$ occurs iff \linebreak $\Delta w_{r',s'} \geq \delta t - 2\epsilon_t$, and $\forall r'_i \in r'$ there exists at least one $s'_j \in s'$ so that $DistS(r'_i, s'_j) \leq \epsilon_{sp}$ and $DistT(r'_i, s'_j) \leq \epsilon_t$, and $\forall s'_j$ there exists at least one $r'_i$ so that $DistS(s'_j, r'_i) \leq \epsilon_{sp}$ and $DistT(s'_j, r'_i) \leq \epsilon_t$. 
\end{definition}

\begin{definition} \label{dfn_4}
\textbf{(Maximally matching subtrajectories)} Given a pair of ``matching'' subtrajectories $(r', s')$ which belong to trajectories $r, s$ respectively, this pair is considered a ``maximal match'' iff $\nexists$ superset $r''$ of $r'$ or $s''$ of $s'$ where the pair $(r'', s')$ or $(r', s'')$ or $(r'', s'')$ would be ``matching''.
\end{definition}

At this point, we should clarify that two trajectories may have more than one ``maximal matches'' (i.e. pairs of subtrajectories). 
Having provided the above background definitions, we can define the subtrajectory join query between two sets of trajectories.

\begin{definition} \label{dfn_5}
\textbf{(Subtrajectory join)} Given two sets of trajectories $R$ and $S$, a spatial threshold $\epsilon_{sp}$, a temporal tolerance $\epsilon_t$ and a time duration $\delta t$, the subtrajectory join query searches for all pairs $(r', s')$, $r' \in r \in R$ and $s' \in s \in S$, which are ``maximally matching'' subtrajectories.
\end{definition}

\subsection{A Closer Look at the Subtrajectory Join Problem} \label{sec_trjoinprop}

An integral part of any algorithm addressing the subtrajectory join query, as defined in Defn.~\ref{dfn_5} above,  is 
to identify all \emph{pairs of joining points} ($r_i$,$s_j$), where $r_i \in r$ and $s_j \in s$, which satisfy the following property: $DistS(r_i, s_j) \leq \epsilon_{sp}$ and $DistT(r_i, s_j) \leq \epsilon_t$. In fact, the set of \emph{joining points} is the outcome of the inner join $R \bowtie S$, where the evaluated join predicates are the ones mentioned above. However, as it will be explained next, these pairs of points do not suffice to return the correct query result. 

Let $\mathcal{A}$ denote the class of correct algorithms for the subtrajectory join problem. 
A naive algorithm $A \in \mathcal{A}$ would require the Cartesian product $R \times S$ to produce the correct result. We claim that $R \times S$ can be represented by three sets of points, the set of \emph{joining points} ($JP$), the \emph{breaking points} ($BP$) and the \emph{non-joining points} ($NJP$). Formally, 
$R \times S = JP \cup BP \cup NJP$.
The definitions of these sets follow, and the discussion is aided by Figure~\ref{fig_intro}(b), which is a variation of Figure~\ref{fig_intro}(a) in order to emphasize the distinction between $JPs$, $BPs$ and $NJPs$.

The first set of points that needs to be identified, besides $JP$, contains all points $r_i \in r \forall r \in R$  that do not ``match'' with any other point in $S$. We call such points as \emph{breaking points}.

\begin{definition} \label{dfn_6}
\textbf{(Breaking points)} A point $r_i \in r \in R$ is a breaking point iff it is not a joining point with any other point $s_j \in S$:

\centerline{$\nexists s_j \in S$: $DistS(r_i, s_j) \leq \epsilon_{sp}$ $\land$ $DistT(r_i, s_j) \leq \epsilon_t$.}
\end{definition}

As it will be shown later, the lack of information about $BPs$ can make an algorithm $A \in \mathcal{A}$ to falsely identify a pair of subtrajectories as ``matching''.
The set of $BP$ along with the set of $JP$ is actually the outcome of the full outer join of $R$ and $S$.  
Figure~\ref{fig_intro}(b) depicts the case where $r_1$ is a breaking point of $r$ ($r_2$, $r_3$ and $r_{10}$ are also breaking points), since it does not ``match'' with any other point of any trajectory. Obviously, breaking points are never reported as part of the answer set and the portion of $r$ that could possibly contribute to the result is subtrajectory $r_{4,9}$.

The last set of points that is necessary to be identified consists of the pairs of points that do not ``match'', coined \emph{non-joining points}, since some of them might indicate the start or the end of ``maximally matching'' subtrajectories.

\begin{definition} \label{dfn_7}
\textbf{(Non-joining points)} A point $r_i \in r \in R$ is a non-joining point w.r.t. $s_j \in s \in S$ iff: (a) $r_i$ and $s_j$ are not breaking points, and (b) $r_i$ is not a joining point with $s_j \in S$:

\centerline{
$r_{i}, s_{j} \notin BP$  $\land$ ($DistS(r_{i}, s_{j}) > \epsilon_{sp}$ $\lor$ $DistT(r_{i}, s_{j}) > \epsilon_t$).
}
\end{definition}

This case is illustrated in Figure~\ref{fig_intro}(b), where $r_5$ is a non-joining point w.r.t. $s_5$. 

Actually, if we remove condition (a) from Definition~\ref{dfn_7} then it is obvious that a \emph{breaking point} $r_i$ is a special case of \emph{non-joining point} where $r_i \in R$ is a \emph{non-joining point} with every other point $\in S$. However, we differentiate \emph{breaking points} from \emph{non-joining points} so as to reduce the amount of information that needs to be kept, i.e. instead of keeping multiple \emph{non-joining points} we only keep one \emph{breaking point}. 
In the section that follows, we investigate the theoretical properties of an efficient algorithm in class  $\mathcal{A}$. 

\subsection{Properties of Subtrajectory Join} \label{sec_trjoinprop2}

In this section, we provide the theoretical properties for designing efficient algorithms for the subtrajectory join problem. The properties shown below essentially determine which pairs of points from the sets $BP$ and $NJP$ are necessary for a correct algorithm in class $\mathcal{A}$.

\begin{lemma} \label{lem_breaking}
The set of breaking points is necessary in order to produce the correct result set for the Subtrajectory Join problem.
\end{lemma}

This result indicates that \emph{breaking points} cannot be ignored by an algorithm, without compromising the correctness of the result. 
The remaining question is whether all \emph{non-joining points} are also necessary.
In the following, we define a subset of \emph{non-joining points} points $sNJP \subseteq NJP$, and show that this subset is actually necessary. 

\begin{definition} \label{dfn_8}
\textbf{(Subset $sNJP$ of non-joining points)} 
A non-joining point $s_j \in S$ w.r.t. a point $r_i \in R$ belongs to $sNJP$, if at least one of its adjacent points $s_{j-1}$ or $s_{j+1}$ is a joining point with any point $r_p \in r$, with $p \neq i$ and $\nexists$ a point $r_q$, with $q \neq i$, such that $DistT(r_q, s_j) \leq DistT(r_i, s_j)$.
\end{definition}

Returning to the example of Figure~\ref{fig_intro}(b), $s_5$ does not ``match'' with any point in $r$, even though all points of $r_{4,8}$ ``match'' with a point in $s_{3,7}$. Again, failure to identify pairs of points such as $(s_{5}, r_{6})$ would result in erroneously identifying larger ``matching'' subtrajectories.

\begin{lemma} \label{lem_nonjoining}
The set $sNJP$ of pairs of non-joining points is necessary in order to produce the correct result set for the Subtrajectory Join problem.
\end{lemma}

In summary, our main finding is that a typical join algorithm that identifies only the set of $JP$ is not enough in order to address the subtrajectory join problem. Additionally to the set of $JP$, an algorithm needs to identify both the set of $BP$ and the subset $sNJP$ during the join processing, in order to ensure correctness. 

\subsection{\prob} \label{sec_challenges}

Given two sets $R$ and $S$ of trajectories, the typical approach for parallel join processing consists of two main phases: (a) \emph{data repartitioning}, in order to create pairs of partitions $R_i \subset R$ and $S_j \subset S$, such that part of the join can be processed using only $R_i$ and $S_j$, and (b) \emph{join processing}, where a join algorithm is performed on partitions $R_i$ and $S_j$.

\begin{problem} \label{pr_1}
\textbf{(\emph{\prob})} Given two distributed sets of trajectories, $R=\cup R_i$ and $S=\cup S_j$, compute the subtrajectory join (Defn.~\ref{dfn_5}) in a parallel manner.
\end{problem}

In this setting, the main challenges are the following: (a) ensure that the created partitions are sufficient to produce parts of the total join without additional data, (b) generate even-sized partitions in order to balance the load fairly to multiple nodes, (c) handle the problem of potential duplicate existence in the join results, which may arise due to the way partitions are created, and (d) process the actual join on the partitions in an efficient way. The first challenge sets the foundations for \emph{parallel processing}, as it identifies pairs of partitions that can be processed together, without any additional data, and produce a subset of the final join result. The second challenge is about \emph{load balancing} and determines the efficiency of parallel processing, which is not straightforward, since processing uneven work units in parallel may lead to sub-optimal performance (as the slowest task will determine the query execution time). The third challenge, labeled \emph{duplicate avoidance}, is about avoiding to generate duplicate results which typically occurs in parallel join processing. Finally, the fourth challenge, labeled \emph{efficient join}, refers to the efficiency of the (centralized) algorithm used to join two partitions.

Clearly, solving the above problem is quite challenging in a distributed setting, as multiple challenges need to be addressed at the same time. In the following sections, we present a well designed solution solution to the \emph{\prob} problem along with two improved versions, following the popular MapReduce paradigm.

\section{The Basic Subtrajectory Join Algorithm} \label{sec_solut}

\subsection{Preliminaries}
One of the prevalent technologies for dealing with Big Data and offline analytics, is the MapReduce programming paradigm~\cite{DBLP_journals/cacm/DeanG10} and its open-source implementation Hadoop~\cite{DBLP_conf/mss/ShvachkoKRC10}. A lot of efforts have been made as far as it concerns join processing through this technology and a survey on limitations of MapReduce/Hadoop, also related to join processing, is conducted in~\cite{DBLP_journals/vldb/DoulkeridisN14}. In more detail, Hadoop is a distributed system created in order to process large volumes of data which are usually stored in the Hadoop Distributed File System (\emph{HDFS}). 
In more detail, when running a MapReduce (MR) job, each \emph{Mapper} processes (in parallel) an input split, which is a logical representation of data. An input split typically consists of a block of data (the default block size is 128MB) but it can be adjusted according to the users' needs by implementing a custom \emph{FileInputFormat} along with the corresponding \emph{FileSplitter} and \emph{RecordReader}. Subsequently, for each record of the split the ``map'' function is applied. The output of the \emph{Map} phase is sorted and grouped by the ``key'' and written to the local disk. Successively, the data is partitioned to \emph{Reducers} based on a partitioning strategy (also known as \emph{shuffling}), and each \emph{Reducer} receives a partition (group) of data and applies the ``reduce'' function to the specific group. Finally, the output of the \emph{Reduce} phase is written to \emph{HDFS}.

\subsection{The DTJb Algorithm}

\setlength\intextsep{0pt}
\begin{figure*} [thb]
  \begin{center}
  \includegraphics[width=1\textwidth]{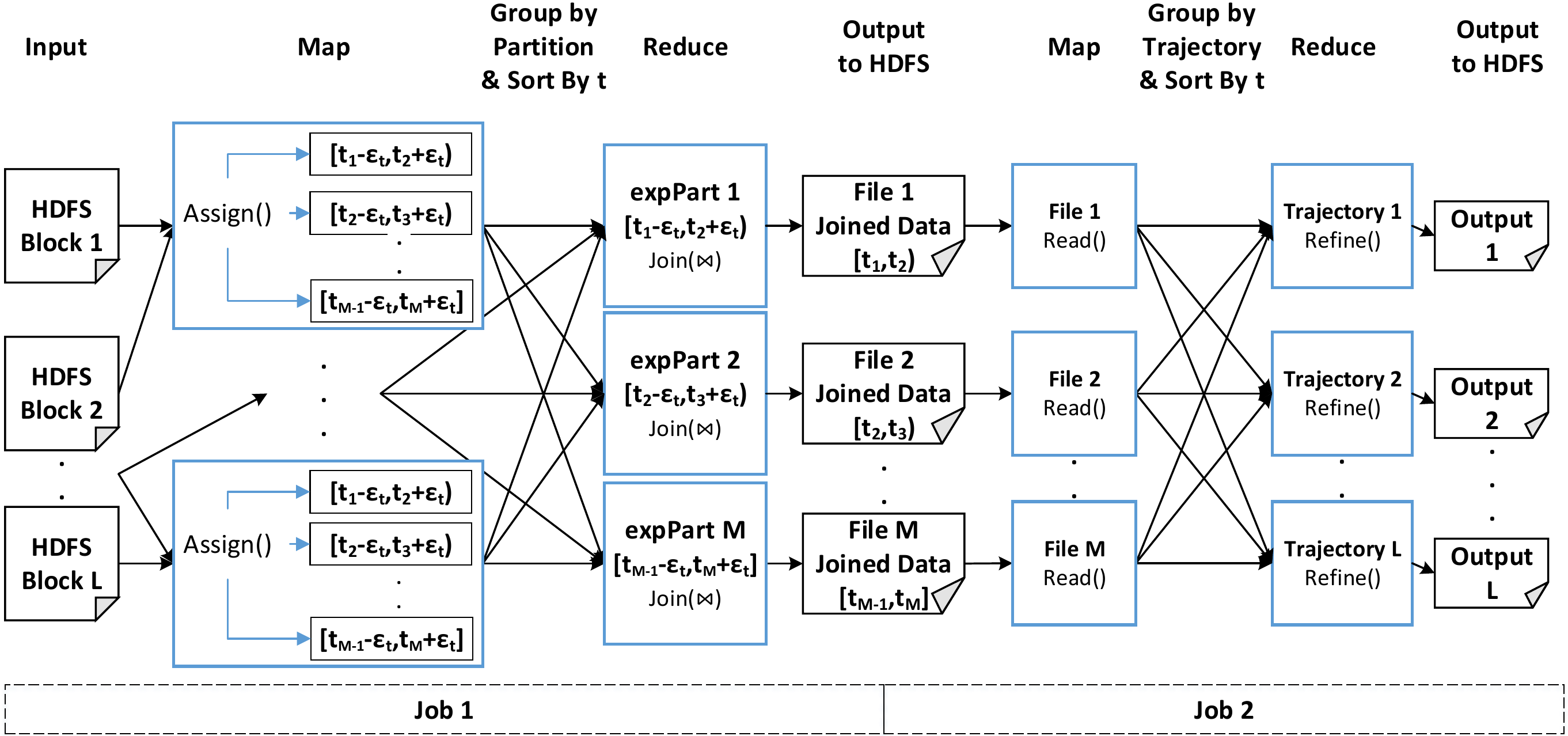}
  \caption{The DTJb algorithm in MapReduce.}
  \label{fig_firstsol}
  \end{center}
\end{figure*}

Our first algorithm, named $\bas$, consists of three phases: (a) the \emph{Partitioning phase}, where input data is read and partitioned, (b) the \emph{Join phase}, where the sets $JP$, $BP$ and $sNJP$ are identified in each partition, and (c) the \emph{Refine phase}, where these sets are grouped  by trajectory and sorted by time in order to identify all the pairs of ``maximally matching'' subtrajectories\footnote{For the sake of simplicity, from now on, we are going to consider the case of self-join. The transition to the problem of joining two relations is straightforward.}.

\subsubsection{Partitioning Phase} \label{sec_partitioningphase}

The first challenge is how to partition the input data in order to satisfy the requirement for \emph{parallel processing}. Partitioning the data into $N$ disjoint temporal partitions $R = \cup_{i=1}^{N}part_i$, where $R$ is the set of trajectories, 
cannot guarantee the correctness during parallel processing, due to the temporal tolerance parameter $\epsilon_t$. 
Hence, we define a partitioning where each $part_i$ is expanded by $\epsilon_t$, thus expanded partitions can be processed independently in parallel. Let $expPart_i$ denote such an expanded partition. 
Processing each $expPart_i$ individually guarantees correctness, but at the cost of having duplicates due to the point replication in temporally overlapping partitions.
To address this \emph{duplication avoidance} challenge, we supplement each point with a flag \emph{partFlag} that indicates whether this point belongs to the original partition (i.e. not expanded by $\epsilon_t$) or not.

\begin{lemma} \label{lem_partitioning1}
An expanded partition $expPart_i$ is sufficient in order to produce the sets of $JP$ and $BP$ for $part_i$
\end{lemma}

Unfortunately, an expanded partition $expPart_i$ is not sufficient in order to produce the set of $sNJP$ since, according to Defn.~\ref{dfn_8}, for each pair $(r_j, s_k)$ that belongs to $NJP$ we need to examine $r_{j-1}$ and $r_{j+1}$, which may span to other partitions. However, the set of $sNJP$ can be identified at the \emph{Refine} phase, where all the pairs concerning a trajectory are grouped together. 

In more detail, we choose to partition the data into uniform temporal partitions, where for each pair of partitions $(part_i, part_j)$, with $i \neq j$ and $i,j \in [1,N]$, it holds that $DistT(t_e^{part_i}, t_s^{part_i}) = DistT(t_e^{part_j}, t_s^{part_j})$. Typically, the duration of a partition is larger than the maximum interval between two consecutive points of any trajectory. As illustrated in Figure~\ref{fig_firstsol}, in the \emph{Map} phase we access each data point and assign it to the expanded partition with which it intersects, essentially applying a temporal range partitioning.
Then, the data is grouped by expanded partition, sorted by time and fed to the \emph{Reduce} phase, where the \emph{Join} procedure takes place. 

\subsubsection{Join Phase} \label{sec_joinphase}

Figure~\ref{fig_firstsol} shows that each \emph{Reducer} task takes as input an expanded partition and performs the \emph{Join} operation. At this point, the duplication avoidance technique is applied, by employing the aforementioned flag and emitting only pairs where at least one point belongs to the original partition. The input of this phase is a set of tuples of the form $\left\langle t, x, y,  trajID, partFlag \right\rangle$. The output of this MR job is a set of (a) $JP$, (b) $BP$ and (c) candidate $sNJP$.

\begin{figure*} [thb]
  \begin{center}
  \includegraphics[width=1\textwidth]{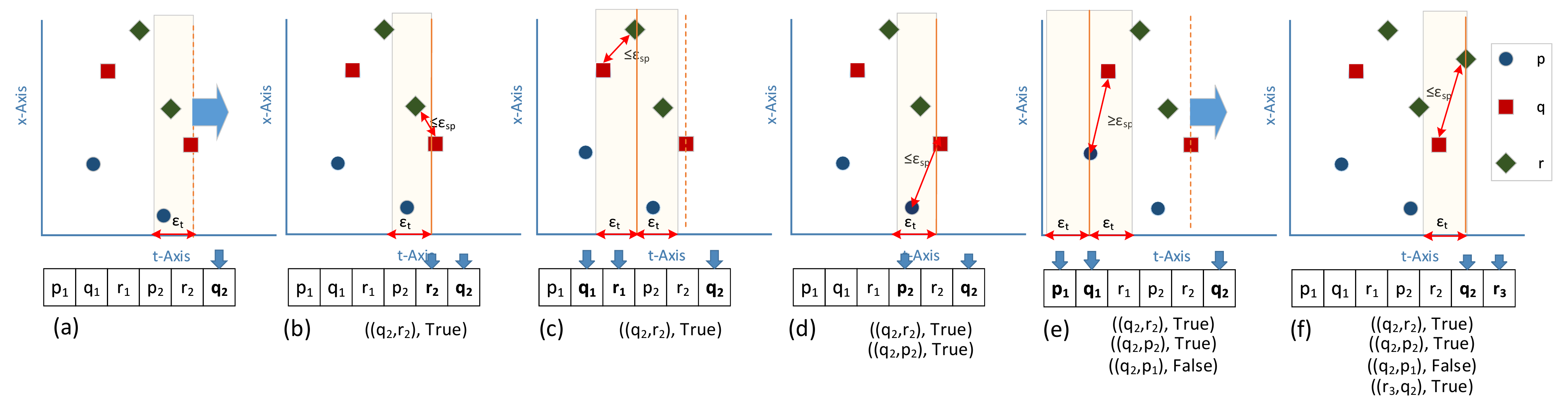}
  \caption{Join phase - The TRJPlaneSweep algorithm.}
  \label{fig_TRJPlaneSweep}
  \end{center}
\end{figure*}

In more detail, we apply a plane sweep technique in order to perform the \emph{Join}, by sweeping the temporal dimension. We choose to employ such a technique due to the fact that is much more efficient than a nested-loop join approach, since our data already arrive sorted by the temporal dimension, as illustrated in Figure~\ref{fig_firstsol}. A typical plane sweep algorithm would emit only the set of $JP$, which is not enough in our case. For this reason, we devised and implemented a modified plane sweep technique, named \emph{TRJPlaneSweep}, depicted in Figure~\ref{fig_TRJPlaneSweep}, which also reports the sets of $BP$ and candidate $sNJP$.

\begin{algorithm}[t]
\caption{Join$(expPart, \epsilon_{sp}, \epsilon_t)$}
\label{alg_join}
\begin{algorithmic}[1]
\STATE{\textbf{Input:} An \emph{expPart}, $\epsilon_{sp}$, $\epsilon_t$}
\STATE{\textbf{Output:} All pairs of $JP$, $BP$ and candidate $sNJP$} 
\FOR{each $point$ $i \in expPart$}
\STATE $D[i] \leftarrow point$
\STATE TRJPlaneSweep($D[], \epsilon_{sp}, \epsilon_t$)
\ENDFOR
\STATE{TreatLastTrPoints()}\label{lin_join0}
\FOR{each $point$ $j \in BP[]$}\label{lin_join1}
\STATE{output($(BP[j], null)$, True)}\label{lin_join2}
\ENDFOR
\end{algorithmic}
\end{algorithm}

\begin{algorithm}[t]
\caption{TRJPlaneSweep($D[], \epsilon_{sp}, \epsilon_t$)}
\label{alg_trjplanesweep}
\begin{algorithmic}[1]
\STATE{\textbf{Input:} $D[]$, $\epsilon_{sp}$, $\epsilon_t$}
\STATE{\textbf{Output:} All pairs of $JP$, $BP$ and candidate $sNJP$} 
\IF{$D[i].partFlag$=True}
\FOR{each element $D[j] \in [D[i].t - \epsilon_t, D[i].t]$} \label{lin_0}
\IF{$DistS(D[i], D[j]) \leq \epsilon_{sp}$}  \label{lin_1start}
\STATE output($(D[i], D[j])$, True) 
\STATE remove $D[i]$ from $BP[]$ \label{lin_remove_B_1}
\IF{$D[j].partFlag$=True}
\STATE output($(D[j], D[i])$, True)
\STATE remove $D[j]$ from $BP[]$ \label{lin_remove_B_2}
\ENDIF      \label{lin_1end}
\STATE $k \leftarrow$ getPrevTrPoint($j, D[]$)   \label{lin_2start}
\IF{FindMatch($D[], i, k, \epsilon_{sp}, \epsilon_t$) = False}
\STATE output($(D[i], D[k])$, False)
\ENDIF
\STATE $k \leftarrow$ getPrevTrPoint($i, D[]$)
\IF{FindMatch($D[], j, k, \epsilon_{sp}, \epsilon_t$)= False}
\IF{$D[j].partFlag$=True}
\STATE output($(D[j], D[k])$, False)  \label{lin_2end}
\ENDIF
\ENDIF
\ENDIF
\ENDFOR
\IF{there is no ``match'' for $D[i]$} \label{lin_3start}
\STATE $BP[] \leftarrow D[i]$
\ENDIF \label{lin_3end}
\ENDIF
\end{algorithmic}
\end{algorithm}

Algorithm~\ref{alg_join} presents how the \emph{Join} processing is performed. Each accessed point is inserted to an array $D$, which contains points sorted in increasing time. After point insertion, (Algorithm~\ref{alg_trjplanesweep} is invoked for the currently accessed point (say $D[i]$) if $D[i]$ belongs to the original partition. \emph{TRJPlaneSweep} examines the previously accessed points for the previous $\epsilon_t$ window (line~\ref{lin_0}).
The role of this function is threefold. First, it identifies \emph{joining points} with $D[i]$, e.g., point $D[j]$, and emits them in the form $((D[i],D[j]),True)$ (lines~\ref{lin_1start}-\ref{lin_1end}). Depending on the outcome of the duplicate avoidance technique, pairs $((D[j],D[i]),True)$ are also output. Second, it discovers points that belong to the candidate $sNJP$ set by  examining whether the previous trajectory point (\emph{getPrevTrPoint})) of $D[j]$ (and $D[i]$), say $D[k]$, is a $NJP$ (\emph{FindMatch}) with each point $\in$ $D[i].trajID$ ($D[j].trajID$, respectively) (lines~\ref{lin_2start}-\ref{lin_2end}). In case such points are identified, they are output with a different flag $((D[i],D[k]),False)$ to differentiate them from $JP$. 
Third, it discovers the points that belong to $BP$. In more detail, in lines~\ref{lin_3start}--\ref{lin_3end}, a \emph{breaking point} $D[i]$ is added to the breaking points set $BP$ and in lines~\ref{lin_remove_B_1} and~\ref{lin_remove_B_2} is removed if a point has a ``match''. The remaining points in $BP$ are reported as breaking points, using the following form: $((D[i],null),True)$ (Algorithm~\ref{alg_join} lines~\ref{lin_join1}--\ref{lin_join2}).

By examining only the previous point of a $JP$ in a trajectory, we might not examine a possible temporary adjacent point that might lie after the last $JP$ of a trajectory in each partition. For this reason, we post-process the last $JP$s in order to check for candidate $sNJP$s by invoking the \emph{TreatLastTrPoints} function (Algorithm~\ref{alg_join} line~\ref{lin_join0}). 

\begin{example}
As illustrated in Figure~\ref{fig_TRJPlaneSweep}(a), we suppose that the current point inserted into $D$ is $q_2$. In Figure~\ref{fig_TRJPlaneSweep}(b), assuming that $DistS(q_2, r_2) \leq \epsilon_{sp}$, we get a ``match'' and pair $((q_2, r_2), True)$ is reported (the symmetric pairs are omitted for simplicity). Subsequently, we need to find the previous point of $r$ and in order to achieve this we should traverse our data backwards until we find it, as presented in Figure~\ref{fig_TRJPlaneSweep}(c). When we find $r_1$, we need to check whether it is a $NJP$ for each point $\in q$, as illustrated in Figure~\ref{fig_TRJPlaneSweep}(c). If there exists a point $\in q$ that ``matches'', in our case $q_1$, nothing is reported and we proceed to examine whether $q_2$ and $p_2$ are $JP$s. If  $DistS(q_2, p_2) \leq \epsilon_{sp}$ then we output the pair $((q_2, p_2), True)$, as shown in Figure~\ref{fig_TRJPlaneSweep}(d). Subsequently, we need to find $p_1$ and check whether it is $NJP$ for each point $\in q$. As depicted in Figure~\ref{fig_TRJPlaneSweep}(e) there is no ``match'' between $p_1$ and any of the points of $q$. For this reason, we report the pair $((q_2,p_1), False)$. The same procedure is continued to the next point inserted to memory as delineated in Figure~\ref{fig_TRJPlaneSweep}(f) until there are no more points inserted.
\end{example}

The complexity of the \emph{Join} procedure is $O(|D| \cdot a \cdot ((1-b) \cdot |D| + b \cdot |D|\cdot(2\cdot(L+2\cdot a \cdot |D|))))$, where $|D|$ is the number of points, $a$ is the selectivity of $\epsilon_t$ and $b$ is the selectivity of $\epsilon_{sp}$. $L$ is the number of points that have to be traversed in order to find the previous point of a specific trajectory.
It is obvious that when $a$ tends to reach 1 the complexity tends to reach $O(|D|^2)$. In the worst case, the complexity can be analogous to $O(|D|^3)$, when both $a$ and $b$ tend to reach 1. However, for a typical analysis task $\epsilon_t$ and $\epsilon_{sp}$ are much smaller than the dataset duration and the dataset diameter respectively. Roughly, we can say that the complexity is $O(a \cdot b \cdot |D|^2)$.

\subsubsection{Refine Phase} \label{sec_refinephase}

\begin{figure} [thb]
  \begin{center}
  \includegraphics[width=.8\textwidth]{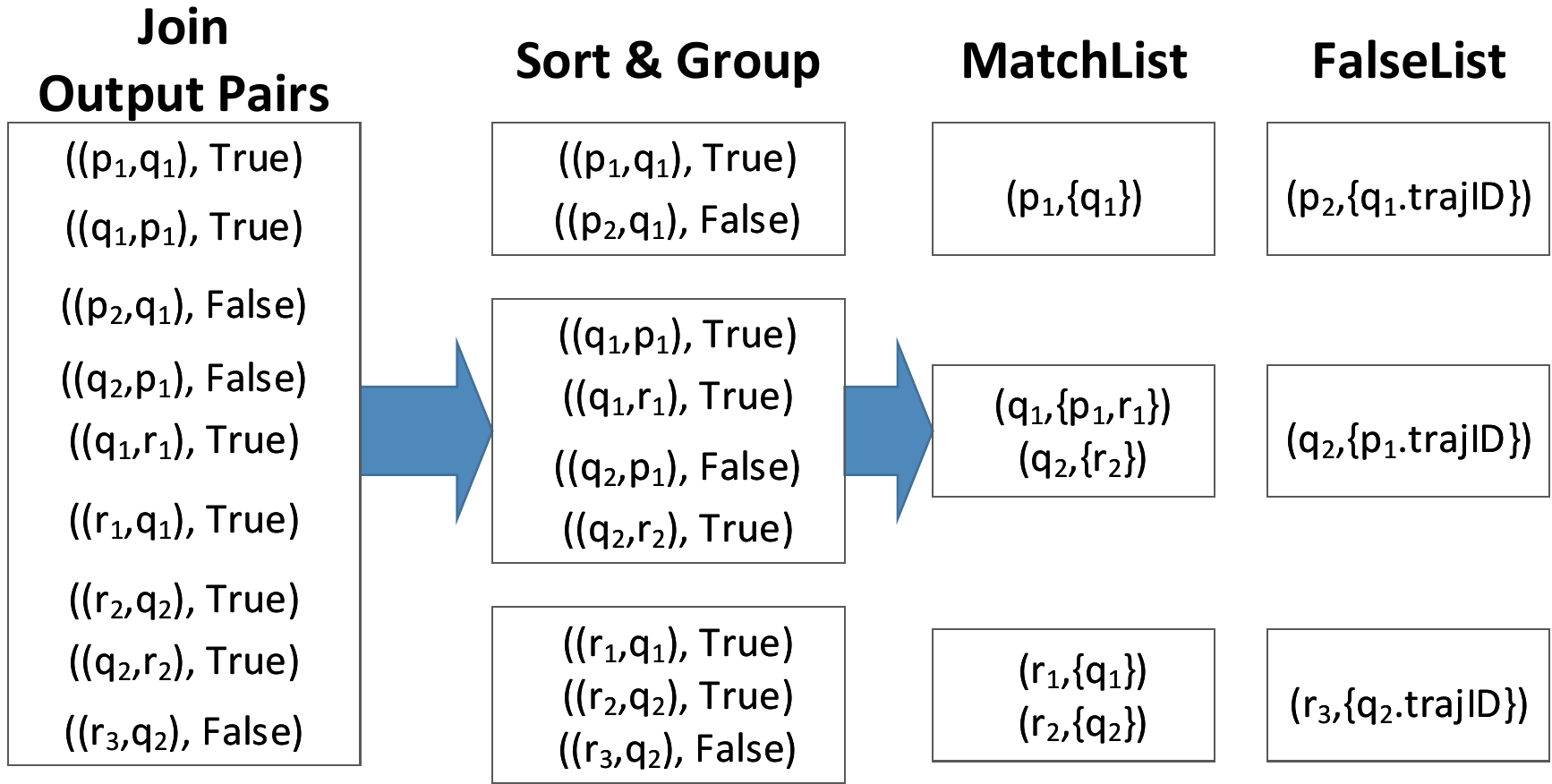}
  \caption{Output of \emph{Join} and input of \emph{Refine} phase.}
  \label{fig_7}
  \end{center}
\end{figure}

The output of the \emph{Join} phase is actually pairs of points. From now on, let us refer to the left point of such a pair as \emph{reference point} and the trajectory that it belongs to, \emph{reference trajectory}.
The \emph{Refine} phase consists of a second MR job that reads the output of the \emph{Join} step and groups points by the \emph{reference trajectory}. Each \emph{Reduce} task receives all pairs of points belonging to a specific trajectory, sorted first by the \emph{reference point's} time and the by the \emph{non-reference trajectory} ID. Figure~\ref{fig_7} shows an example where the output pairs of points from the \emph{Join} step are grouped, sorted and fed as input to three \emph{Reduce} tasks (for trajectories $p$, $q$, and $r$ respectively). The general idea here is to scan the set of $JP$ in a sliding window fashion so as to identify ``maximally'' matching subtrajectories and at the same time ``consult'' the sets of $BP$ and $sNJP$ in order to avoid false identifications, as described in Section~\ref{sec_trjoinprop2}.

Hence, each \emph{Reducer} accesses all the pairs of a \emph{reference trajectory} (say $p$) sorted by time, i.e., $\{p_1,p_2,\dots,p_n\}$. 
Algorithm~\ref{alg_refine} describes the pseudo-code of the \emph{Refine} phase which aims to identify all the ``maximally matching'' pairs of subtrajectories of $p$ with other subtrajectories of any trajectory $x$ ($x \neq p$). For each accessed pair (($p_i,x_j$),\emph{flag}), the algorithm assigns it in one of the two structures that it maintains: the \emph{MatchList} and the \emph{FalseList}. All $JP$ and $BP$ will be kept in the \emph{MatchList}, whereas the candidate $sNJP$ is kept in the \emph{FalseList} (lines~\ref{lin_updMatch1}--\ref{lin_updMatch2}). Again, this is more clearly depicted in the example of Figure~\ref{fig_7}. Also, notice that for each \emph{reference point} in the \emph{MatchList}, we maintain a list of points sorted by trajectory ID.

\begin{lemma} \label{lem_partitioning2}
The set of candidate $sNJP$ is sufficient so as to identify the set of $sNJP$ at the \emph{Refine} phase.
\end{lemma}

\begin{algorithm}[t]
\caption{Refine($\delta t$, $\epsilon_t$)}
\label{alg_refine}
\begin{algorithmic}[1]
\STATE{\textbf{Input:} Pairs of points (($p_i,x_j$),\emph{flag}) for a given trajectory $p$, sorted by time}
\STATE{\textbf{Output:} Result of \emph{\prob} for $p$} 
\FOR{each pair of points (($p_i,x_j$),\emph{flag})}
\IF{($p_i$ is encountered for the first time)}
\IF{DistT(\emph{MatchList}.lastEntry, \emph{MatchList}.firstEntry)  $\geq \delta t$}  \label{lin_1}
\STATE \emph{resultT} $\leftarrow$ intersect lists in \emph{MatchList} and exclude \emph{FalseList} \label{lin_2}
\STATE \emph{resultF} $\leftarrow$ apply sliding window of $\delta t$ to \emph{resultT} \label{lin_3}
\STATE \emph{resultFinal} $\leftarrow$ \emph{resultFinal} $\bigcup$ \emph{resultF} \label{lin_4}
\STATE remove \emph{MatchList}.firstEntry \label{lin_5}
\ENDIF
\ENDIF
\IF{(\emph{flag} = True)} \label{lin_updMatch1}
\STATE addToMatchList($p_i,x_j$)
\ELSE
\STATE addToFalseList($p_i,x_j$)
\ENDIF
\ENDFOR  \label{lin_updMatch2}
\STATE output(\emph{resultFinal}) \
\end{algorithmic}
\end{algorithm}

\begin{figure} [thb]
  \begin{center}
  \includegraphics[width=0.8\textwidth]{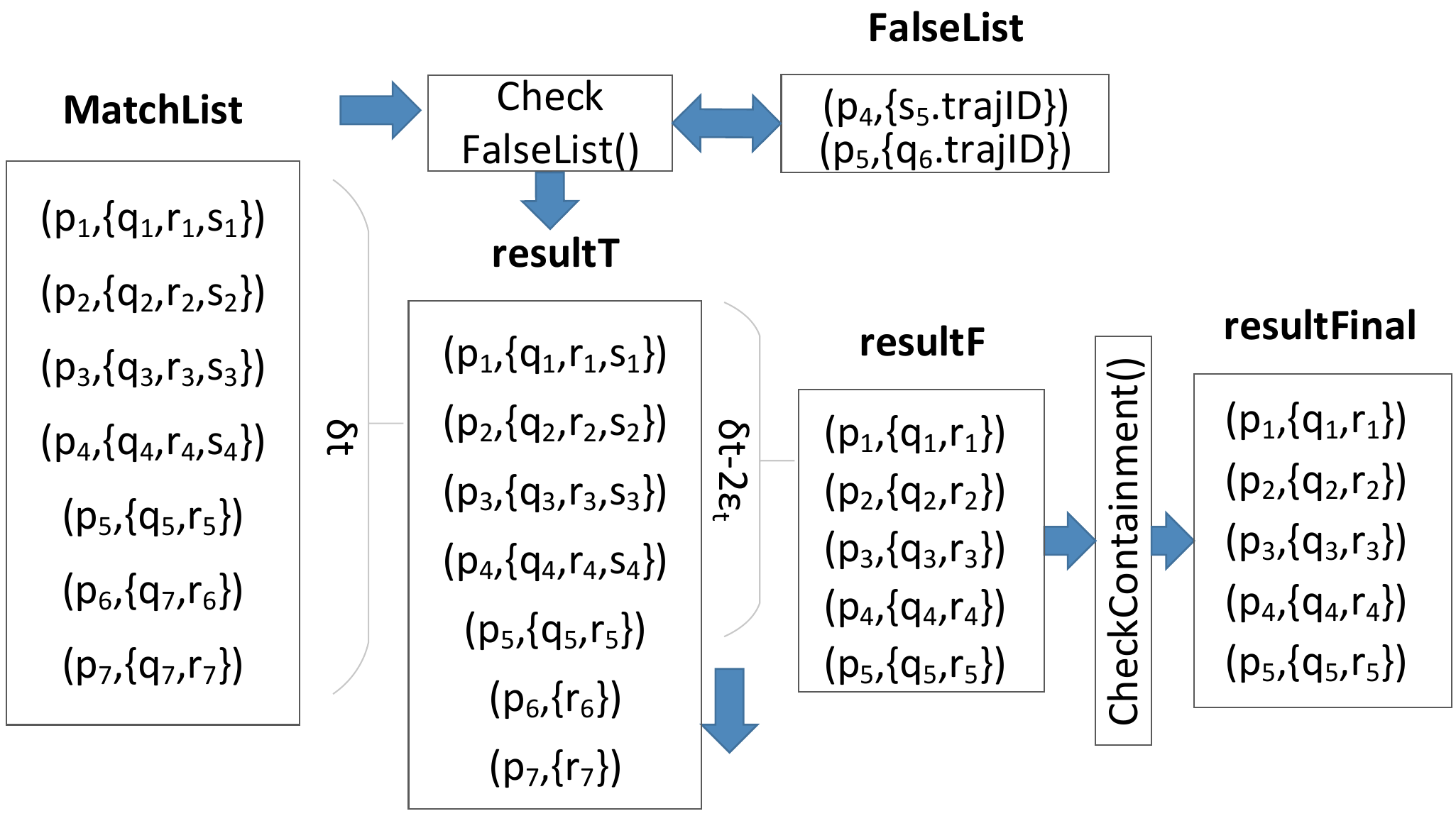}
  \caption{Refine procedure.}
  \label{fig_refine}
  \end{center}
\end{figure}

The algorithm proceeds as follows: as soon as all pairs of points of a specific \emph{reference point} $p_i$ have been accessed, it initiates processing on the \emph{MatchList}. The processing takes place only if the first and last point of $p$ in \emph{MatchList} have temporal distance greater than or equal to $\delta t$ (line~\ref{lin_1}). 
The processing essentially identifies points of other trajectories that join with points of $p$ in the whole temporal window.
This is performed by intersecting the lists in \emph{MatchList} and excluding points existing in the \emph{FalseList} (line~\ref{lin_2}). List intersection is efficiently performed in linear time to the length of the lists, since the lists are sorted by trajectory ID. Figure~\ref{fig_refine} depicts the result of this processing as \emph{resultT}.

Subsequently, the points in \emph{resultT} are processed as follows. We start from the first point and take into consideration all points with temporal distance at most $\delta t - 2 \epsilon_t$ from the first point. From this set of points, we derive the subtrajectories that ``match'' for the entire $\delta t - 2 \epsilon_t$ window, and insert them in \emph{resultF} (line~\ref{lin_3}).
The temporary results of the \emph{resultF} structure are added to the final result structure \emph{resultFinal}, if not already contained in it (line~\ref{lin_4}). Then, a new set of points is considered, of temporal distance at most $\delta t - 2 \epsilon_t$ from the second point of \emph{resultT} and the process is repeated, similarly to a sliding a window of duration $\delta t - 2 \epsilon_t$ on \emph{resultT}.
In the end, the first entry of the \emph{MatchList} ($p_1,\{q_1,r_1,s_1\}$) is removed (line~\ref{lin_5}), as all potential results containing $p_1$ have already been produced.
The algorithm terminates when the entire trajectory is traversed, the \emph{resultFinal} is returned and each element of this list is emitted. 

\begin{example}
Figure~\ref{fig_refine} presents a working example of the Refine algorithm given the specific \emph{MatchList} and \emph{FalseList} of trajectory $p$.Assuming that $DistT(p_{1}.t, p_{7}.t) \geq \delta t$, we intersect all the lists contained in the specific window of the \emph{MatchList} and we pass the result to \emph{resultT}. In this way, the list of the last entry of \emph{resultT} will contain only the points that belong to the subtrajectories that move ``close'' enough with $p$ for the whole $\delta t$ window. During list intersection, we take into account the \emph{FalseList} structure in order to deal with points that belong to $sNJP$. Specifically, even though for each $p_{i}$, with $i \in$ [1,7] $\exists$ a ``match'' with $q$, however $q_{6}$ has no ``match'' with $p$, as depicted in the \emph{FalseList}. For this reason, $q$ should be excluded from \emph{resultT} after $p_{5}$. Then, a sliding $\delta t - 2 \epsilon_t$ window is created that traverses \emph{resultT}, and for each such window we intersect all lists and the result is stored in  \emph{resultF}. For the first  $\delta t - 2 \epsilon_t$ window, as depicted in Figure~\ref{fig_refine}, subtrajectories $r_{1,5}$ and $q_{1,5}$ are identified.  The reason for this is to discover the subtrajectories that move ``close'' enough, with $p$ for the whole $\delta t - 2 \epsilon_t$ window. Subsequently, before proceeding to the next  $\delta t$ window, the contents of \emph{resultF} are inserted to the final result, if not already contained.
\end{example}

The complexity of the \emph{Refine} procedure is $O(T\cdot SW\cdot dt\cdot l)$, where $T$ is the average number of points in a trajectory, $SW$ is the number of points contained in the $\delta t$ window, $dt$ is the number of points contained in the $\delta t - 2 \epsilon_t$ window and $l$ is the size of the list. The complexity, here, clearly depends on the average number of points per trajectory, the $\epsilon_t$ and $\delta t$ parameter, and the number of pairs emitted by the join phase which in turn depends on $\epsilon_t$ and $\epsilon_{sp}$.

\section{Subtrajectory Join with Repartitioning} \label{sec_algor}

Even though the $\bas$ algorithm provides a correct solution to the \emph{\prob} problem, it has some limitations. In particular, it does not address the \emph{load balancing} challenge, since it does not handle the case of temporally skewed data. Also, due to the two chained MR jobs, the intermediate output of the first job is written to HDFS and must be read again by the second job, which imposes a significant overhead as its size is comparable and can be even bigger than the original dataset. 

Motivated by these limitations, we propose an improved two-step algorithm ($\rep$), which consists of the \emph{repartitioning} and the \emph{query} step. Each step is implemented as a MR job. However, the repartitioning step is considered a preprocessing step, since it is performed once and is independent of the actual parameters of our problem, namely $\epsilon_{sp}$, $\epsilon_t$, and $\delta t$.

\begin{figure*} [thb]
  \begin{center}
  \includegraphics[width=1\textwidth]{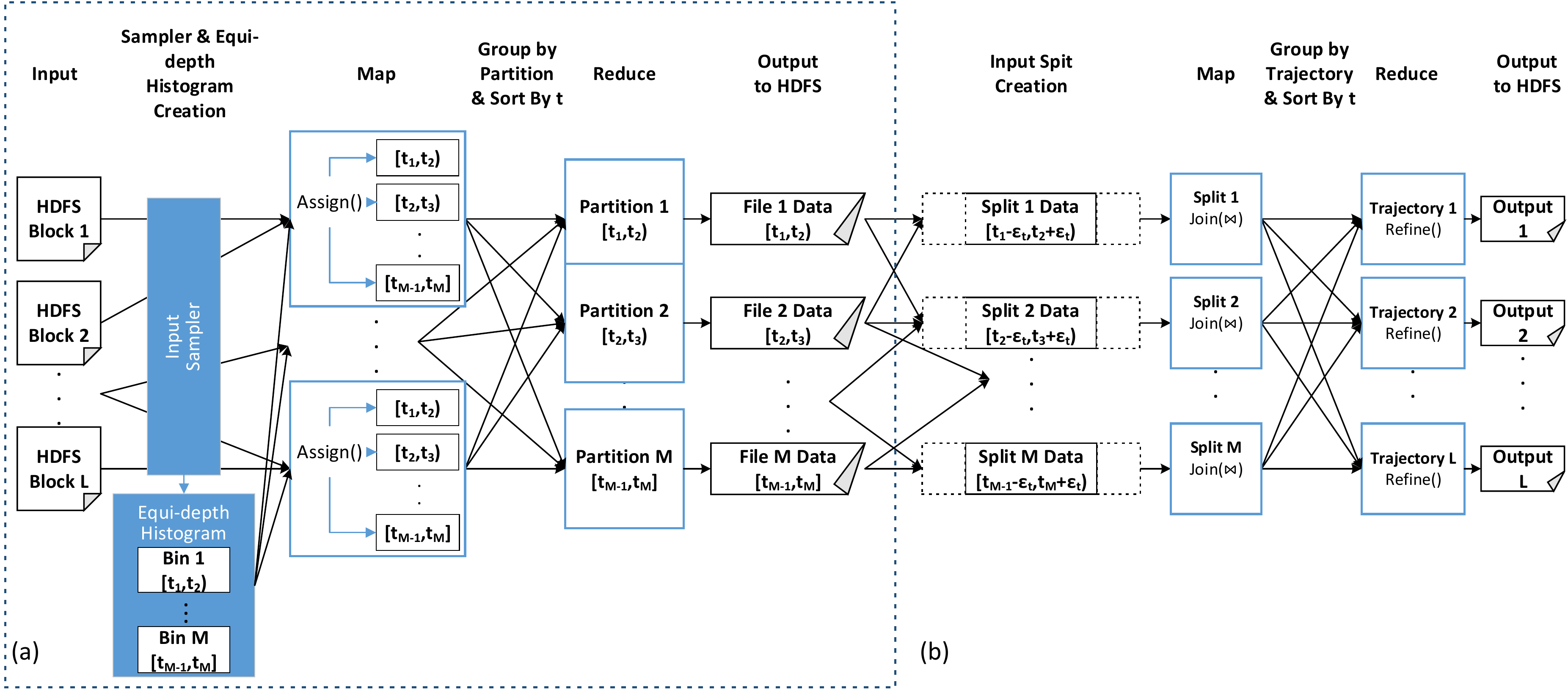}
  \caption{The DTJb algorithm in MapReduce: (a) Repartitioning step and (b) Query step.}
  \label{fig_RJR}
  \end{center}
\end{figure*}

\subsection{Repartitioning} \label{sec_repart}

The aim of the repartitioning step is to split the input dataset in $M$ equi-sized, temporally-sorted partitions (files), which are going to be used as input for the join algorithm. This is essential for two reasons: (a) it will provide the basis for load balancing, by addressing the issue of temporal skewness in the input data, and (b) it will result in temporal collocation of data, thus drastically reducing processing and network communication costs.

The repartitioning step is performed by means of a MR job as follows. We sample the input data, using Hadoop's \emph{InputSampler}, and construct an equi-depth histogram on the temporal dimension. The histogram contains $M$ equi-sized bins, i.e. the numbers of points in any two bins are equal, where the borders of each bin correspond to a temporal interval $[t_i,t_j)$.

The equi-depth histogram is exploited by the \emph{Map} phase in order to assign each incoming data object in the corresponding histogram bin, based on the value of its temporal dimension. Each ``map'' function outputs each data object using as key a value $[1,M]$ that corresponds to the bin that the object belongs to. During shuffling, all data objects that belong to a specific bin are going to be sorted in time and will be collected by a single ``reduce'' function (thus having $M$ ``reduce'' functions). As a result, each ``reduce'' function writes an output file to \emph{HDFS} that contains all data objects in a specific temporal interval $[t_i,t_j)$ sorted by increasing time. A graphical view of the MR job is provided in Figure~\ref{fig_RJR}(a). 

A subtle issue is how to determine the number $M$ of bins (and, consequently, output files). A small value of $M$, smaller than the number of nodes in the cluster, would be opposed to the collocation property because data would have to be transferred through the network. On the other hand, a large value of $M$ would result to many small files, smaller than the \emph{HDFS} block size, and would lead to inefficient use of resources as well as increasing the management cost of these \emph{HDFS} files. A good compromise is to have files of equal size to the \emph{HDFS} block. Hence, the number of files can be calculated as $M = \lceil \frac{InputTotalSize}{hdfs block size} \rceil$. Collocation can be further improved by placing temporally adjacent files to the same nodes. This can be achieved by grouping together $k$ consecutive files, where $k = \lceil \frac{M}{N} \rceil$, with $N$ being the number of nodes, and using their group id as the partition key. 

\subsection{The DTJr Algorithm}

In order to minimize the I/O cost, the MR job that implements the proposed algorithm performs the \emph{Join} procedure in the \emph{Map} phase, and the \emph{Refine} in the \emph{Reduce} phase. To achieve this, we need to provide to a \emph{Map} task as input, a data partition that contains all necessary data in order to perform part of the \emph{Join} procedure \emph{independently} from other \emph{Map} tasks. Thus, an \emph{HDFS} block produced by the repartitioning phase is expanded with additional points that exist at time (+/-)$\epsilon_t$, and this is the process of \emph{InputSplits} creation.
In this way, points are duplicated to other \emph{HDFS} blocks, which means that the same point may be output by two different \emph{Map} tasks.
To avoid this pitfall, a different \emph{duplicate avoidance mechanism} is introduced which practically determines that a point is going to be output only by a single \emph{Map} task; the \emph{Map} task processing the \emph{HDFS} block where the point belongs to.

As already mentioned, each data partition (\emph{InputSplit}) that is fed to a \emph{Map} task should contain all the data needed to perform the join of points for the specific partition, i.e. data for the period $[t_s^{part} - \epsilon_t, t_s^{part} + \epsilon_t]$.
However, an output file produced by the repartitioning step is not sufficient due to the temporal tolerance $\epsilon_t$, thus we need to augment these output files with extra data points, so that they form independent data partitions.
At technical level, we devised and implemented a new \emph{FileInputFormat} called \emph{BloatFileInputFormat}, along with the corresponding \emph{FileSplitter} and \emph{RecordReader}, which selectively combines different files in order to create splits that carry all the necessary data points. Furthermore, during the creation of input splits we augment (as metadata) each split with the starting and ending time of the original partition of each split, termed $t_s^{base}$ and $t_e^{base}$. The utility is to provide us with a simple way to perform duplicate avoidance at the \emph{Join} phase.

Figure~\ref{fig_RJR}(b) shows that each \emph{Map} task takes as input a split and performs the join at the level of point for a specific data partition. The input of this phase is a set of tuples of the form $\left\langle t, x, y,  trajID \right\rangle$ sorted in ascending time $t$ order. 
Since the data are already sorted w.r.t. the temporal dimension, we can apply the \emph{Join} procedure, presented in Section~\ref{sec_joinphase}.
The output of the \emph{Map} phase will be the $JP$, $BP$ and $sNJP$ sets. Finally, the \emph{Refine} procedure presented in Section~\ref{sec_refinephase} can be performed at the \emph{Reduce} phase.

\section{Index-based Subtrajectory Join with Repartitioning} \label{sec_index}

The \emph{Join} step of the previous algorithms is common and operates on the array $D$ that contains temporally sorted points. However, it 
can be improved in two ways. First, by employing spatial filtering in order to avoid attempting to join points that are far away. Second, by having an index structure that given a point $p_i$ can efficiently locate the (temporally) previous point $p_{i-1}$ of $p$. 
Motivated by these observations, we devised and implemented an indexing scheme in order to speed up the processing of the join.

\subsection{Indexing Scheme}

\begin{figure} [thb]
  \begin{center}
  \includegraphics[width=0.8\textwidth]{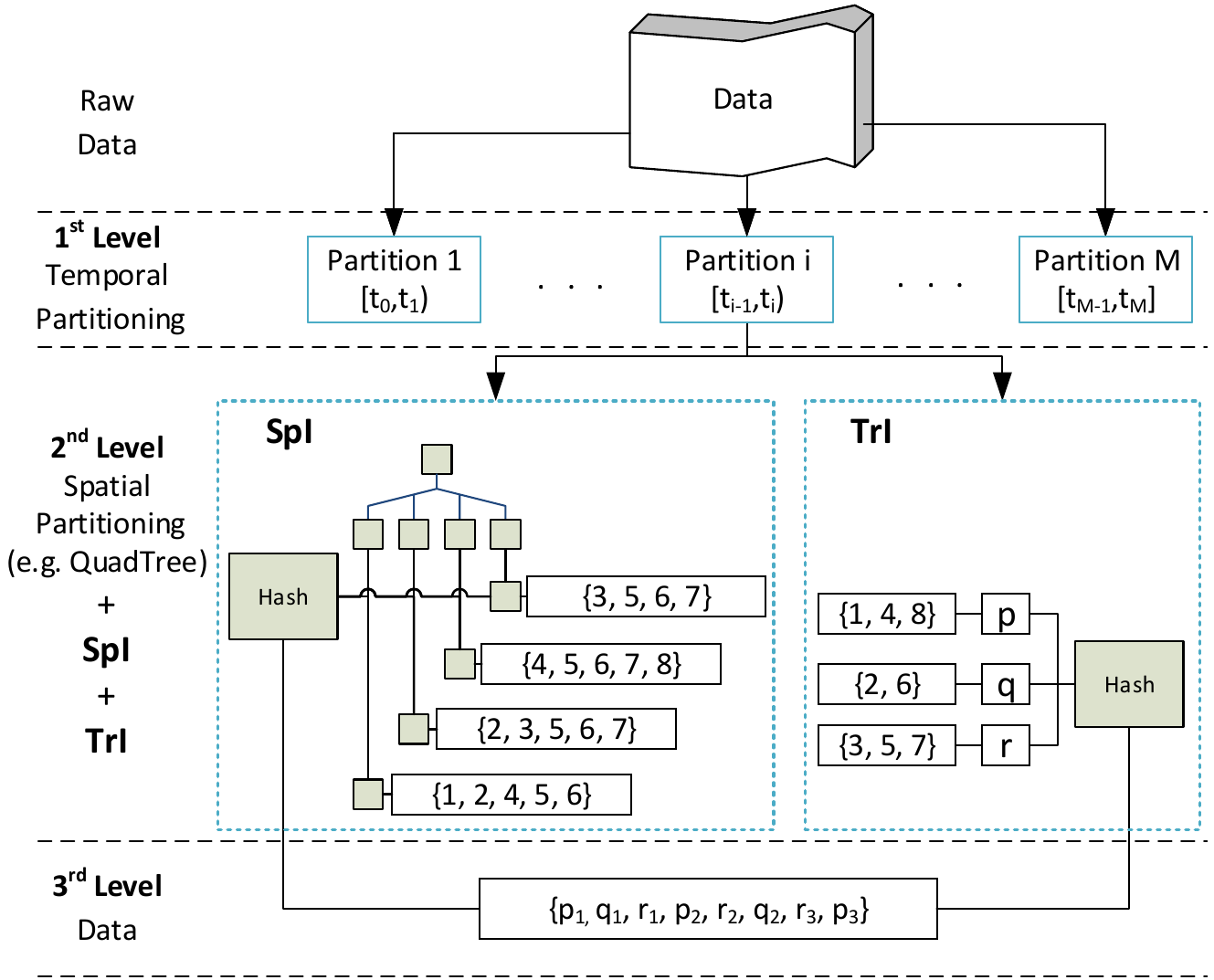}
  \caption{Indexing Scheme of $\ind$ algorithm}
  \label{fig_9}
  \end{center}
\end{figure}

As illustrated in Figure~\ref{fig_9}, this scheme consists of 3 levels. We already covered the first level in Section~\ref{sec_repart}, where the initial data are partitioned to equi-sized temporal partitions (Section~\ref{sec_algor}). At the second level, we partition the space. In order to have load balanced partitions we utilize the spatial partitioning provided by QuadTrees. More specifically, an ``empty'' QuadTree is created once, by sampling the original data, as in ~\cite{DBLP_conf/icde/EldawyM15}, and is written to \emph{HDFS}. It is important to mention here that the QuadTree contains only the spatial partitions and not the actual points. Then, when a new query is posed, the QuadTree is loaded into Hadoop's \emph{distributed cache} in order to be accessible by all the nodes. Moreover, at the same level, we employ two indexes. The first index is a spatial index (SpI) which enables pruning of points based on their spatial distance, thus decreasing significantly the number of points that need to be examined within the $\epsilon_t$ window. The second index is an index that keeps track of the representation of each individual trajectory within the temporally sorted structure $D$ (TrI), thus providing an efficient way to access the previous trajectory point. The two indexes are created gradually, as the data are read from HDFS. Finally, at the third level, we have the temporally sorted data that correspond to the specific temporal partition.

\subsubsection{Spatial Index (SpI)}

The spatial index, called SpI, utilizes a given space partitioning, in our case QuadTrees. For each spatial partition of the QuadTree, SpI keeps a temporally sorted array where each entry is the position of a point that is contained in the given partition expanded by $\epsilon_{sp}$. SpI is implemented as a HashMap with key the partition id and value the sorted array. Thus, a partition can be accessed in O(1), while a point in a partition can be accessed in $O(logP_i)$, where $P_i$ here is the number of points in the corresponding sorted array.
The construction of SpI has $O(|{D}|\cdot h)$ complexity, where $|{D}|$ is the number of points in the specific temporal partition and $h$ is the height of the QuadTree, since for each point we need to traverse the QuadTree in order to find out in which expanded partition it is contained. Note that each point is enriched with the id of its original (i.e. not expanded) spatial partition, thus consisting of $\left\langle  trajID, x, y, t, PartitionID \right\rangle$.

\subsubsection{Trajectory Index (TrI)}

The TrI index keeps track of each individual trajectory within $D$. TrI is also implemented as a HashMap with key the trajectory id. For each trajectory, the value is a temporally sorted array, where each entry corresponds to a point of a trajectory, and the value of the entry is an integer indicating the point's position in $D$. Thus, a trajectory point can be efficiently accessed in $O(logT)$, where $T$ is the number of points of a trajectory. To exemplify, the first element of the array holds the position of the first point of the trajectory inside $D$ and so on. The construction of this index has $O(T)$ time complexity since the data is already sorted in time.

\subsection{The DTJi Algorithm}

\begin{algorithm}[t]
\caption{Join$^I$ (Split, $\epsilon_{sp}$, $\epsilon_t$, $t_s^{base}$, $t_e^{base}$)}
\label{alg_JoinI}
\begin{algorithmic}[1]
\STATE{\textbf{Input:} A split, $\epsilon_{sp}$, $\epsilon_t$, $t_s^{base}$, $t_e^{base}$}
\STATE{\textbf{Output:} All pairs of $JP$, $BP$ and candidate $sNJP$} 
\STATE $QT\leftarrow LoadQuadTree()$ \label{lin_JoinI1}
\FOR{each $point$ $i \in Split$} \label{lin_JoinI2}
\IF{$point.t \in [t_s^{base} - \epsilon_t, t_e^{base} + \epsilon_t]$}
\STATE $D[i], TrI, SpI \leftarrow point$ 
\ENDIF
\STATE TRJPlaneSweep$^I$($D[], TrI, SpI, \epsilon_{sp}, \epsilon_t, t_s^{base}, t_e^{base}$) \label{lin_JoinI3}
\ENDFOR
\STATE{TreatLastTrPoints()}
\FOR{each $point$ $j \in BP[]$}
\STATE{output($(BP[j], null)$, True)}
\ENDFOR
\end{algorithmic}
\end{algorithm}

\begin{algorithm}[t]
\caption{TRJPlaneSweep$^I$($D[], TrI, SpI, \epsilon_{sp}, \epsilon_t, t_s^{base}, t_e^{base}$)}
\label{alg_trjplanesweepI}
\begin{algorithmic}[1]
\STATE{\textbf{Input:} $D[]$, $\epsilon_{sp}$, $\epsilon_t$, $t_s^{base}$, $t_e^{base}$}
\STATE{\textbf{Output:} All pairs of $JP$, $BP$ and candidate $sNJP$} 
\IF{DuplCheck($D[i].t, t_s^{base}, t_e^{base}$)=True}
\FOR{each element $D[j]$ returned by getCandidatePoint($i, SpI, D[]$)} \label{lin_trjplanesweepI1}
\IF{DistS($D[i], D[j]$) $\leq \epsilon_{sp}$}
\STATE output($(D[i], D[j])$, True)
\STATE remove $D[i]$ from $BP[]$
\IF{DuplCheck($D[j].t, t_s^{base}, t_e^{base}$)=True}
\STATE output($(D[j], D[i])$, True)
\STATE remove $D[j]$ from $BP[]$
\ENDIF
\STATE $k \leftarrow$ getPrevTrPoint$^I$($j, D[], TrI$); \label{lin_trjplanesweepI2}
\IF{FindMatch$^I$($D[], i, k, \epsilon_{sp}, \epsilon_t, TrI$)= False} \label{lin_trjplanesweepI4}
\STATE output(($D[i], D[k]$), False)
\ENDIF
\STATE $k \leftarrow$ getPrevTrPoint$^I$($i, D[], TrI$); \label{lin_trjplanesweepI3}
\IF{FindMatch$^I$($D[], j, k, \epsilon_{sp}, \epsilon_t, TrI$) = False} \label{lin_trjplanesweepI5}
\IF{DuplCheck($D[j].t, t_s^{base}, t_e^{base}$)=True}
\STATE output(($D[j], D[k]$), False)
\ENDIF
\ENDIF
\ENDIF
\ENDFOR
\IF{there is no ``match'' for $D[i]$}
\STATE $BP[] \leftarrow D[i]$
\ENDIF
\ENDIF
\end{algorithmic}
\end{algorithm}

Having these two indexes at hand we can utilize them in order to perform the join operation in an efficient way. Algorithm~\ref{alg_JoinI}, presents the index-enhanced plane sweep procedure. Initially, the QuadTree is loaded into memory from the \emph{distributed cache} (line~\ref{lin_JoinI1}) and then, each accessed point is inserted not only to an array $D$, which contains points sorted in increasing time, but also to the SpI and TrI indexes. Finally, the TRJPlaneSweep$^I$() algorithm is invoked for each accessed point (lines~\ref{lin_JoinI2}--\ref{lin_JoinI3}).

Algorithm~\ref{alg_trjplanesweepI}, presents the TRJPlaneSweep$^I$() algorithm. Here, given a point $p_i \in p$, instead of scanning the whole $\epsilon_t$ window before it, in order to find ``matches'', we perform a search in SpI and get only the points that belong to the same partition as $p_i$ by invoking the getCandidatePoint() method (line~\ref{lin_trjplanesweepI1}). The partition id is retrieved in $O(1)$ and then binary search is performed in the temporally sorted list of points in order to find the position of $p_i$ inside it. Having that, we can get the previous element, which will be the previous point in time that lies within the same partition, and check if the temporal and spatial constraint are satisfied. If they are satisfied, we have a ``match'', we proceed to the previous element of SpI and so on and so forth. Assuming that we have a ``match'' with $q_j$ that belongs to trajectory $q$ we need to find the previous point of $q$. This is achieved by invoking getPrevTrPoint$^I$, which performs a search in TrI in order to retrieve in O(1) the entry of $q$ (lines~\ref{lin_trjplanesweepI2}, \ref{lin_trjplanesweepI3}). Then, by performing binary search in the temporally sorted list, we can find the position of $q_j$ and can easily get $q_{j-1}$. Having that, we need to find if it ``matches'' with any point that belongs to $p$. Here, instead of scanning the whole $2 \epsilon_t$ window of $q_{j-1}$ in order to check for ``matches'' with $p$, we perform a search in TrI in order to get the points of $p$ that exist ``close'' to the time of $q_{j-1}$ (lines~\ref{lin_trjplanesweepI4}, \ref{lin_trjplanesweepI5}). Then, if the spatial and temporal constraints are satisfied we have a ``match'' and the \emph{FindMatch$^I$()} method returns True. Otherwise, the whole procedure continues, until the temporal constraint is not satisfied anymore.

The complexity of the index-based solution is $O(|D| \cdot h \cdot (log_2 P_i \cdot a \cdot P_i((1-b) \cdot P_i + b \cdot P_i \cdot (2 \cdot (log_2 T + (log_2 T + a \cdot T)))))))$, with $|D|$ being the number of points, $h$ the height of the QuadTree, $a$ and $b$ the selectivity of $\epsilon_t$ and $\epsilon_{sp}$ respectively. $P_i$ is the number of points within the $i$-th partition expanded by $\epsilon_{sp}$, where $P_i \ll |D|$, and $T$ is the number points per trajectory. In the worst case, where $a$ and $b$ tend to 1, the complexity can reach $O(|D|\cdot log_2 P_i\cdot P_{i}^{2})$. However, again this only occurs for values of $\epsilon_t$ and $\epsilon_{sp}$ that are comparable to the dataset's duration and diameter respectively. Roughly speaking, the complexity drops to $O(|D|\cdot (log_2 P_i \cdot a \cdot b \cdot P_{i}^{2}))$, which clearly shows the benefit attained when employing the proposed indexing scheme. 

\section{Experimental Study} \label{sec_exper}

In this section, we provide our experimental study on the comparative performance of the three variations of our solution, namely (1) $\bas$ that uses two MR jobs (Section~\ref{sec_solut}), (2) $\rep$ that employs repartitioning and a single job to perform the join (Section~\ref{sec_algor}), and (3) $\ind$ that additionally uses the SpI and TrI indexes for more efficient join processing (Section~\ref{sec_index}). Furthermore, we compare our solution with the work presented in~\cite{DBLP_conf/cluster/ZhangHLWX09}.

The experiments were conducted in a 49 node Hadoop 2.7.2 cluster,  provided by \emph{~okeanos}\footnote{\url{https://okeanos.grnet.gr/home/}}, an IAAS service for the Greek Research and Academic Community. 
The master node consists of 8 CPU cores, 8 GB of RAM and 60 GB of HDD while each slave node is comprised of 4 CPU cores, 4 GB of RAM and 60 GB of HDD. Our configuration enables each slave node to launch 4 containers, thus resulting that at a given time the cluster can run up to 192 jobs (\emph{Map} or \emph{Reduce}). 
The real dataset employed for our experiments is IMIS\footnote{IMIS dataset has been kindly provided by IMIS Hellas for research and educational purposes. It is available for downloading at \url{http://chorochronos.datastories.org}}, which consists of 699,031 trajectories of ships moving in the Eastern Mediterranean for a period of 3 years. This dataset contains approximately 1.5 billion records, 56GB in total size. 

Our experimental methodology is as follows: Initially, we verify the scalability of our algorithms by varying (a) the dataset size, and (b) the number of cluster nodes (Section~\ref{sec_scalab}). Then, we examine the benefits of the repartitioning step as well as the associated cost (Section~\ref{sec_loadbal}). Successively, we compare our solution with the work presented in~\cite{DBLP_conf/cluster/ZhangHLWX09} (Section~\ref{sec_dtjvssjmr}). Subsequently, we perform a sensitivity analysis in order to evaluate the effect of different parameters to our algorithms (Section~\ref{sec_sensit}). Finally, we perform a set of experiments so as to examine the creation time and the size of the proposed indexes w.r.t. to varying the number of spatial partitions and $\epsilon_{sp}$ (Section~\ref{sec_idx}).

Table~\ref{tab_params} shows the experimental setting, where we vary the following parameters: $\epsilon_t$, $\epsilon_{sp}$, $\delta t$, the maximum number of points per cell, and the number of cluster nodes, which are the main parameters affecting the performance of our algorithms. 
In more detail, the values of $\epsilon_{sp}$ were calculated as a percentage of the diameter of the smallest cell produced by the QuadTree, whereas the maximum number of points per cell is calculated as a percentage over the total population.

\begin{table}[thb]
\begin{small}
\begin{center}
\caption{Parameters and default values (in bold)} \label{tab_params}
\begin{tabular}{|l|l|} \hline
 \textbf{Parameter}&\textbf{Values}\\ \hline \hline
$\epsilon_t$ (in minutes)	& 10, 15, \textbf{20}, 25, 30 \\ \hline
$\epsilon_{sp}$ (\%) & 10\%, 20\%, \textbf{30\%}, 40\%, 50\% \\ \hline
$\delta t$ (in minutes)	& 10, 15, \textbf{20}, 25, 30 \\ \hline
max \# of points per cell (\%) &	1\%, 2\%, \textbf{3\%}, 4\%, 5\% \\ \hline
\# of Nodes &	12, 24, 36, \textbf{48} \\ \hline
\end{tabular}
\end{center}
\end{small}
\end{table}

\subsection{Scalability} \label{sec_scalab}

\begin{figure*} [thb]

    \begin{minipage}[t]{1\linewidth}
      \centering
      \begin{minipage}[t]{\linewidth}
      \centering
      (a)\hspace*{-0.5cm}\includegraphics[width=.5\linewidth]{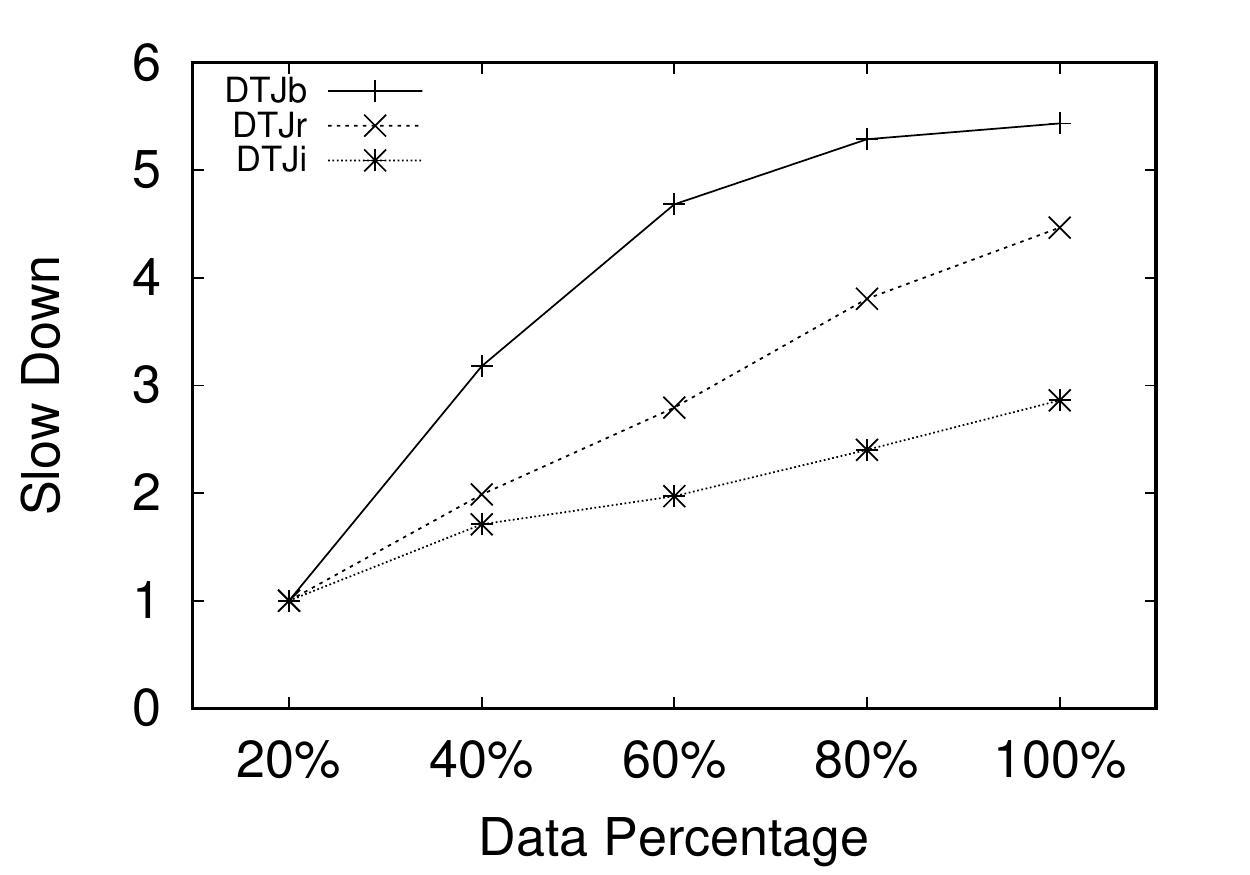}
      (b)\hspace*{-0.35cm}\includegraphics[width=.5\linewidth]{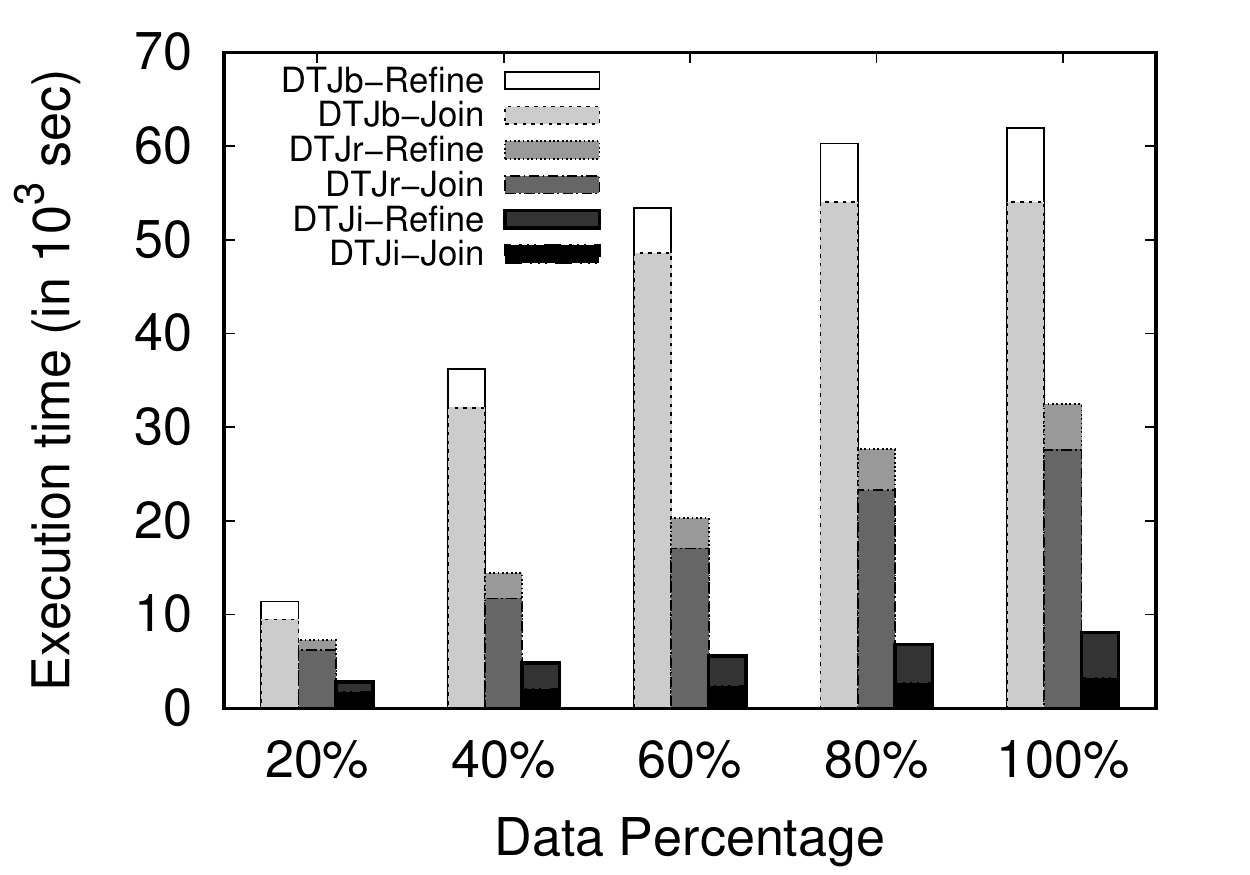}
      \hspace*{-1cm}
      \end{minipage}\par
      \begin{minipage}[t]{\linewidth}
      \centering
      (c)\hspace*{-0.5cm}\includegraphics[width=.5\linewidth]{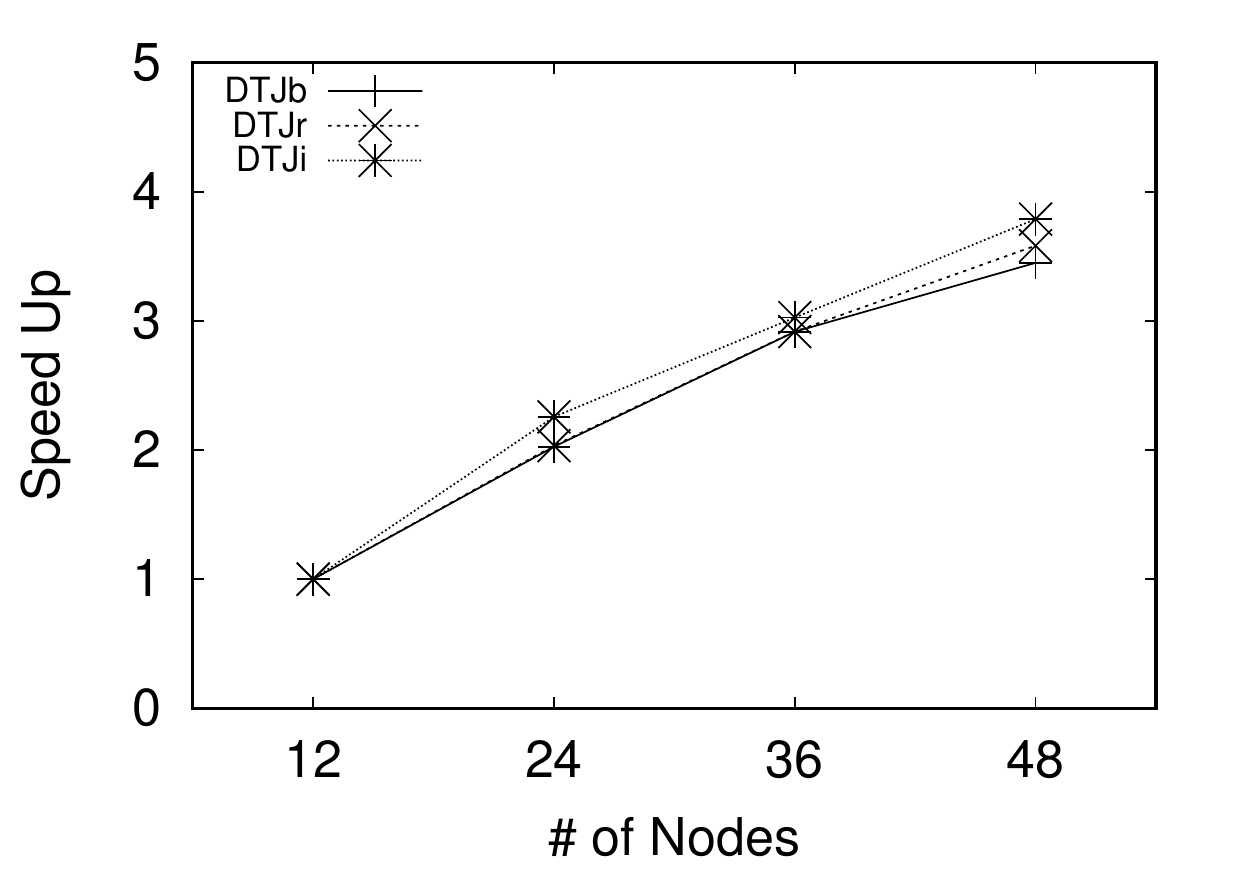}
      (d)\hspace*{-0.35cm}\includegraphics[width=.5\linewidth]{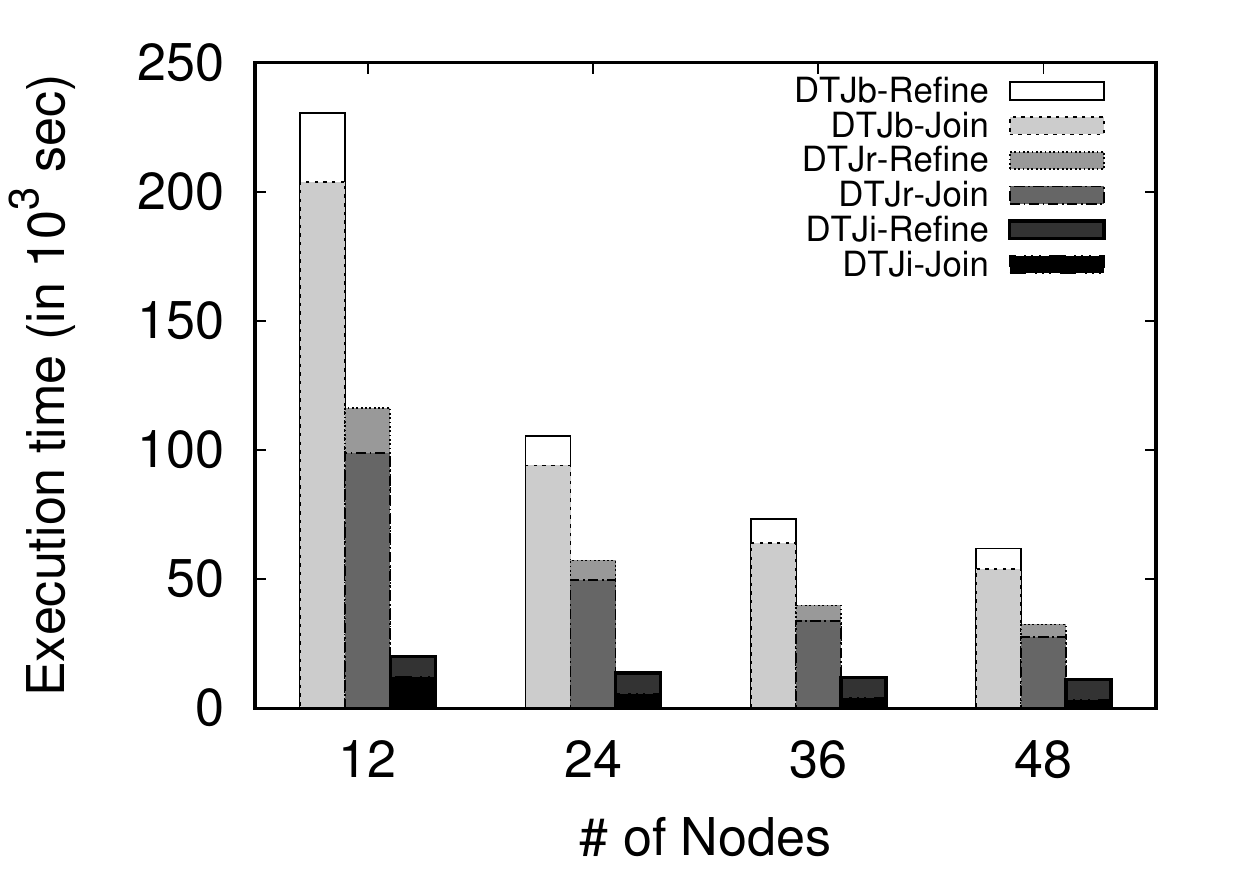}
      \hspace*{-1cm}
      \end{minipage}
      \caption{Scalability analysis varying (a),(b) the size of the dataset and (c),(d) the number of nodes}
      \label{fig_Scalability}
    \end{minipage}%

\end{figure*}

Initially, we vary the size of our dataset and measure the execution time of our algorithms. To study the effect of dataset size, we created 4 portions (20\%, 40\%, 60\%, 80\%) of the original dataset. As the dataset size increases and the number of nodes remains the same, it is expected that the execution time will increase. In order to measure this, for each portion  $D_i$ of the dataset with $i \in [1,5]$, we calculate $Slow Down = \frac{T_{D_i}}{T_{D_1}}$, where $T_{D_1}$ is the execution time of the first portion (i.e. 20\%) and $T_{D_i}$ the execution time of the current one.  
As shown in Figure~\ref{fig_Scalability}(a), as the size of the dataset increases, the $\rep$ linear and the $\ind$ appears to have linear behaviour, with $\ind$ presenting better scalability. On the other hand, $\bas$ appears to have a somehow ``abnormal'' behaviour. This can be justified if we study Figure~\ref{fig_rep_lb}, which presents the standard deviation of the different portions of the dataset. In fact, we can observe that $\bas$ in Figure~\ref{fig_Scalability}(a) is affected by how imbalanced is the partitioning in each portion of the dataset, as depicted in $\bas$ Figure~\ref{fig_rep_lb}.
To investigate further the performance of the different algorithms, we measure separately the execution time of the  \emph{Join} and  \emph{Refine} phases for all algorithms. Concerning the  \emph{Join} phase, as illustrated in Figure~\ref{fig_Scalability}(b), $\ind$ outperforms $\bas$ by 16$\times$ and the $\rep$ by almost one order of magnitude. Regarding the  \emph{Refine} phase, as depicted in Figure~\ref{fig_Scalability}(b), $\rep$ and $\ind$, perform exactly the same, as anticipated, since they use an identical algorithm. Instead, $\bas$ performs worse due to the fact that the  \emph{Refine} phase is implemented as a second MR job, which means that the output of the  \emph{Join} phase, which is typically several times larger than the input data, needs to be read from \emph{HDFS} and get sorted, grouped and shuffled to the \emph{Reduce} tasks.

Subsequently, we keep the size of the dataset fixed (at 100\%) and vary the number of nodes. As the number of nodes increase and the dataset size remains the same, it is expected that the execution time will decrease. In order to measure this, for each portion  $N_i$ of the dataset with $i \in [1,5]$, we calculate $Speed Up = \frac{T_{N_i}}{T_{N_1}}$, where $T_{N_1}$ is the execution time when using the minimum number of nodes (i.e. 12) and $T_{N_i}$ the execution time of the current one. In this experiment, as illustrated in Figure~\ref{fig_Scalability}(c) and (d), we observe that all three approaches present linear scaling, with $\ind$ demonstrating slightly better scalability. The reason why the behaviour is different here, is that in this experiment the dataset that was employed was fixed (100\%). This means that, despite the fact that we vary the number of nodes, the effect of the different algorithms over the data is the same. On the contrary, when we increase the amount of data, as already shown in Figure~\ref{fig_Scalability}(b), we can see than the performance of $\bas$ is affected by the skewness of the different portion of the dataset that were used. As depicted in Figure~\ref{fig_rep_lb}, we can observe that the standard deviation of $\bas$ affect significantly $\bas$ in Figure~\ref{fig_Scalability}(a) and (b).

\subsection{Repartitioning and Load Balancing} \label{sec_loadbal}

\begin{figure*} [thb]
    \begin{minipage}[t]{1\linewidth}
      \begin{minipage}[t]{\linewidth}
          \centering
          (a)\hspace*{-0.5cm}\includegraphics[width=.5\linewidth]{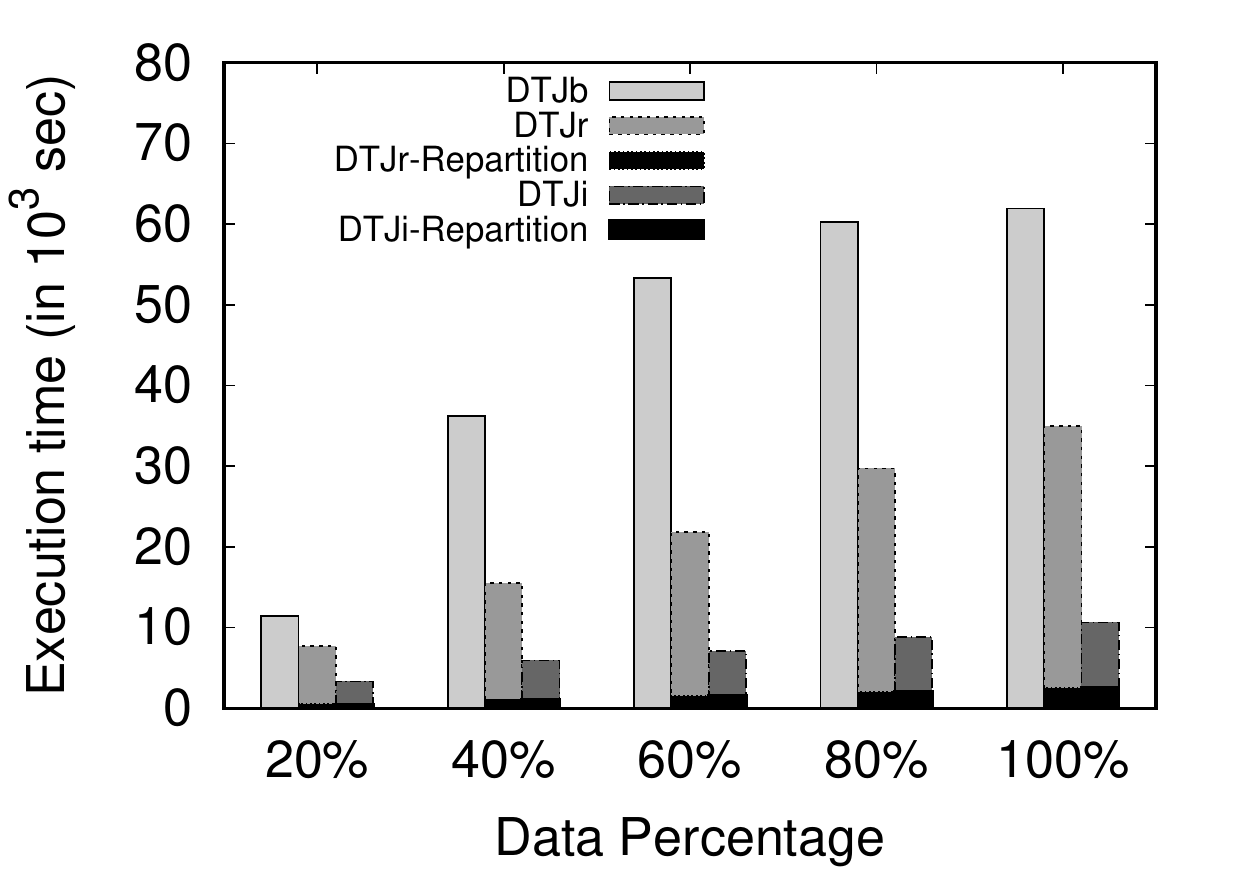}
          (b)\hspace*{-0.35cm}\includegraphics[width=.5\linewidth]{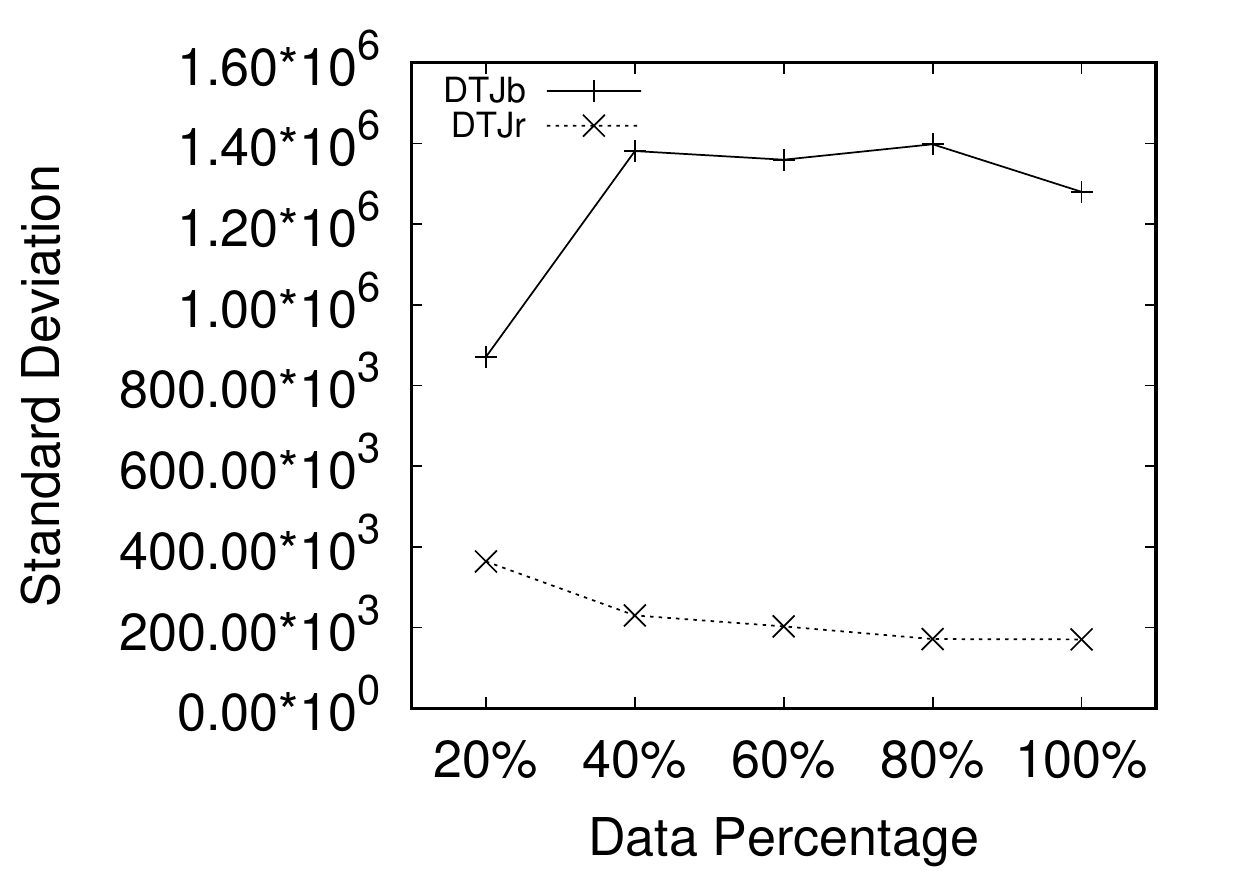}
        \hspace*{-1cm}
      \end{minipage}
      \caption{(a) Repartitioning cost and (b) Load balancing}
      \label{fig_rep_lb}
    \end{minipage}
\end{figure*}

In this set of experiments, we evaluate the cost of the repartitioning step employed by $\rep$ and $\ind$. In Figure~\ref{fig_rep_lb}(a), we compare $\bas$ (which does not use this step) against $\rep$ and $\ind$ after including in the latter two algorithms the time needed for repartitioning. The result shows that even for a single query both algorithms outperform $\bas$. Obviously, for multiple queries with different query parameters ($\epsilon_t$, $\epsilon_{sp}$, $\delta t$), the gain is multiplied, since the repartitioning cost needs to be paid only once, before processing the first query. This experiment justifies the use of the repartitioning step, while demonstrating its low overhead in the case of a single query, which in the case of multiple queries becomes negligible.

In order to quantify whether the work allocation of the \emph{Join} is balanced to the different parallel tasks, we compare the input size of the \emph{Join} phase of $\bas$, against the \emph{Join} phase of $\rep$. In Figure~\ref{fig_rep_lb}(b), we report the standard deviation of the size of input data for the various tasks. Smaller values of the standard deviation, indicate that the different tasks are assigned with similar-sized input data, thus the load is more fairly balanced.
$\rep$ demonstrates significantly lower standard deviation, approximately one order of magnitude, than $\bas$. This also partly justifies the overall better performance of $\rep$ illustrated in Figure~\ref{fig_Scalability}(b) and Figure~\ref{fig_Scalability}(d).

\subsection{Comparative Evaluation} \label{sec_dtjvssjmr}

As already mentioned, the problem of \emph{\prob} has not been addressed yet in the literature and it is not straightforward (if and) how state of the art solutions to trajectory similarity search and trajectory join can be adapted to solve the problem. However, if we utilize only a specific instance of our problem, when $\delta{t} = 0$, then we only need to identify the set of $JP$ during the \emph{Join} phase. Based on this observation, we select to compare with the work presented in ~\cite{DBLP_conf/cluster/ZhangHLWX09}, called $SJMR$, a state of the art MapReduce-based spatial join algorithm, which is able to identify efficiently the set of $JP$ that will be passed to the \emph{Refine} procedure and produce the desired result. The reason why $SJMR$ was chosen is that it is a generic solution which could form the basis for any distributed spatial join algorithm and thus required the minimum amount of modifications so as to match with our problem specification.

More specifically, $SJMR$ repartitions the data at the Map phase and Joins them at the Reduce phase by performing a plane sweep join. For the sake of comparison, we modified $SJMR$ by injecting time as a third dimension and introducing parameters $\epsilon_{sp}$ and $\epsilon_{t}$.  In more detail. at the Map phase the spatiotemporal space is divided to tiles using a fine grained grid. Then, each data point is expanded by $\epsilon_{sp}$ and $\epsilon_{t}$ and is assigned to the tiles with which it intersects. Subsequently, the tiles are mapped to partitions using the method described in ~\cite{DBLP_conf/cluster/ZhangHLWX09}. At the Reduce phase, the points are grouped by partition and sorted by one of the dimensions (we chose the temporal dimension so as to be aligned with our solution). Finally, we sweep through the time dimension and report the set of $JP$.

\begin{figure} [thb]
  \begin{center}
  \includegraphics[width=.5\linewidth]{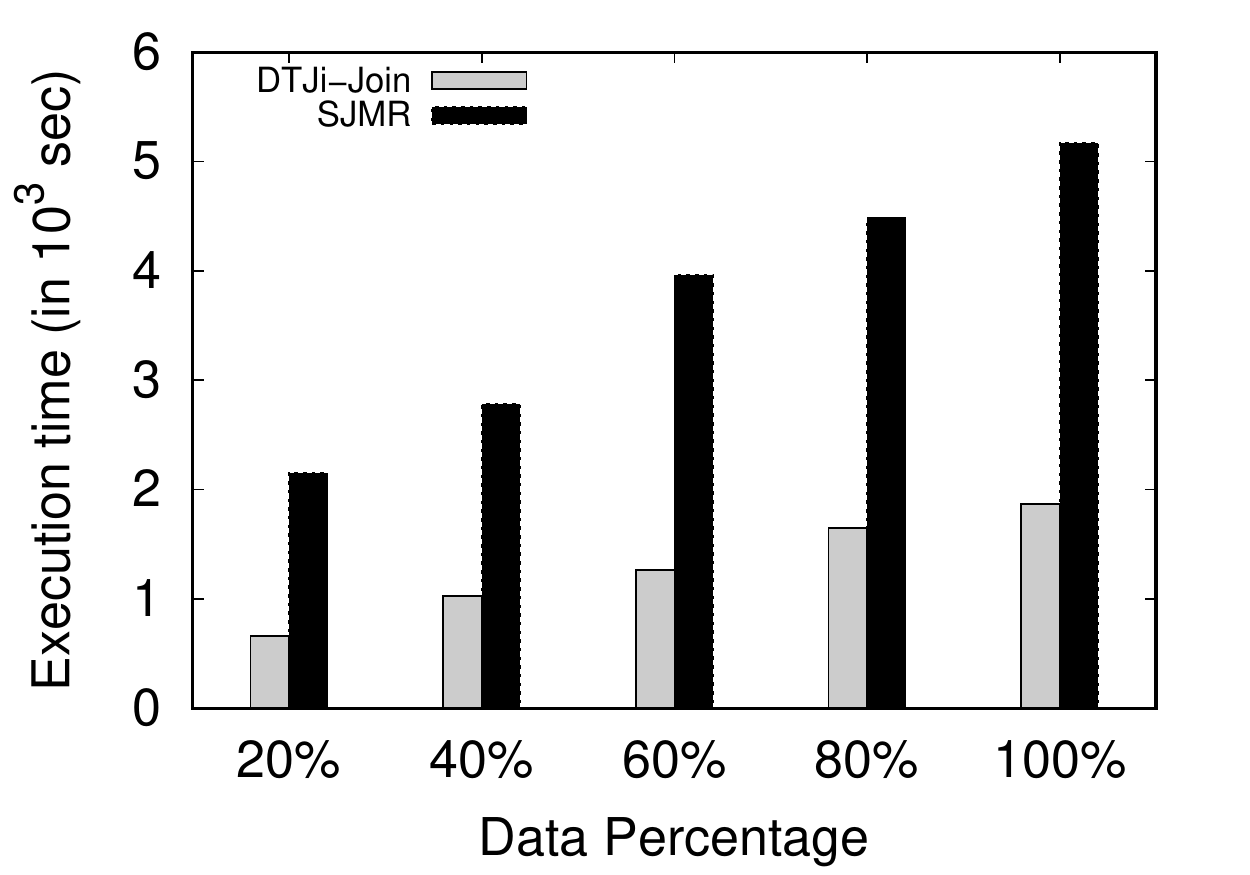}
  \caption{Comparative evaluation between $\ind$ and $SJMR$}
  \label{fig_dtj-vs-sjmr}
  \end{center}
\end{figure}

So, in this set of experiments we compare $\ind$-Join, which outperforms $\bas$-Join and $\rep$-Join, with $SJMR$. In more detail, we vary the size of our dataset and measure the execution time of the two algorithms. The results, as illustrated in Figure~\ref{fig_dtj-vs-sjmr} show that $\ind$-Join not only performs significantly better than $SJMR$ but more importantly, the gain of $\ind$-Join over $SJMR$ increases for larger data sets. The reason for this behaviour lies mainly due to the utilization of the indexing structure of $\ind$ (~\cite{DBLP_conf/cluster/ZhangHLWX09} uses no indexes) and the fact that $\ind$-Join is a Map-only job where the repartitioning cost is ``paid'' only once (as a preprocessing step), unlike $SJMR$, where this cost is ``paid'' every time at the Map phase, as explained earlier.

\subsection{Sensitivity Analysis} \label{sec_sensit}

In the following experiment, we perform a sensitivity analysis of algorithms $\rep$ and $\ind$. We exclude $\bas$ from this set of experiments, as it consistently performs significantly worse than the other two algorithms.

\begin{figure*} [thb]
  \begin{minipage}[t]{1\linewidth}
  \centering
  (a)\includegraphics[width=.45\linewidth]{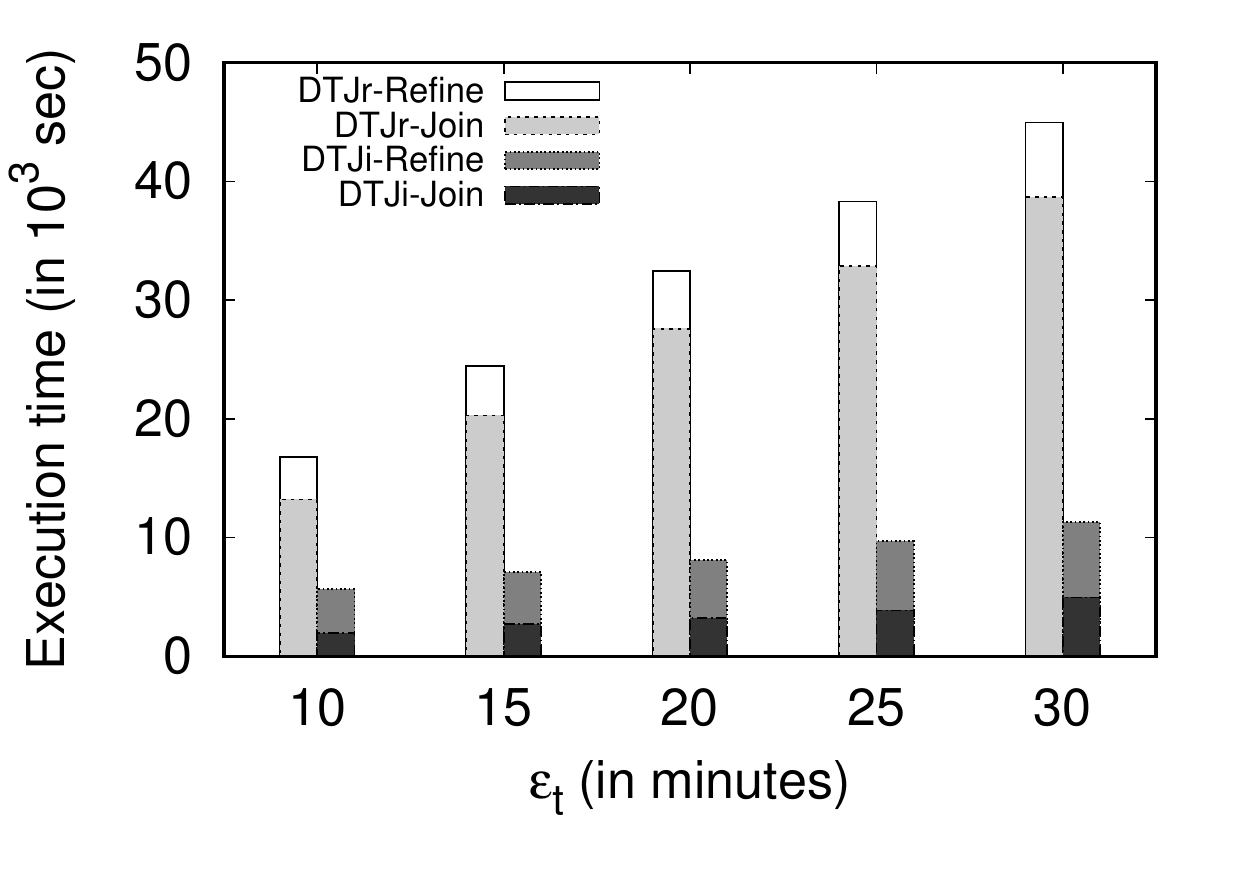}
  (b)\includegraphics[width=.45\linewidth]{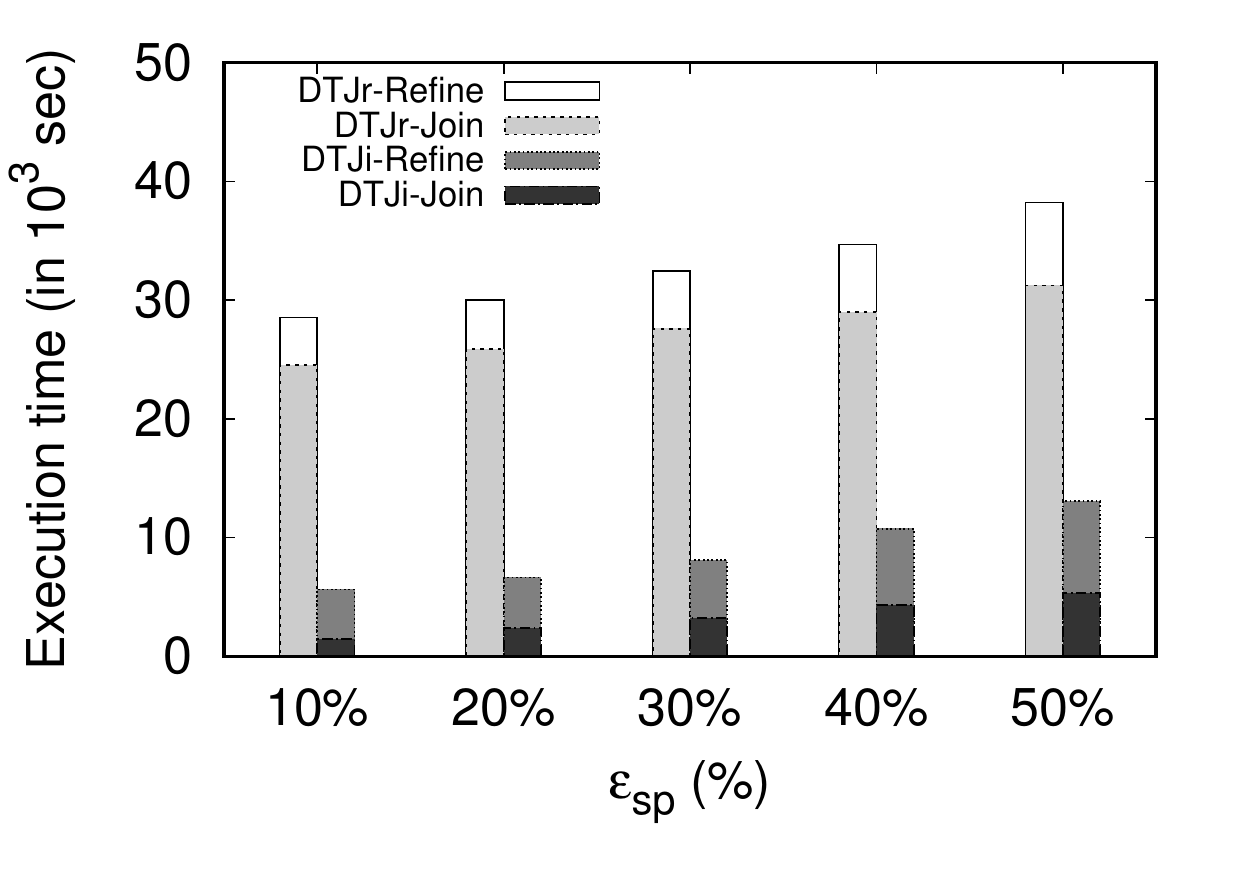}
  (c)\includegraphics[width=.45\linewidth]{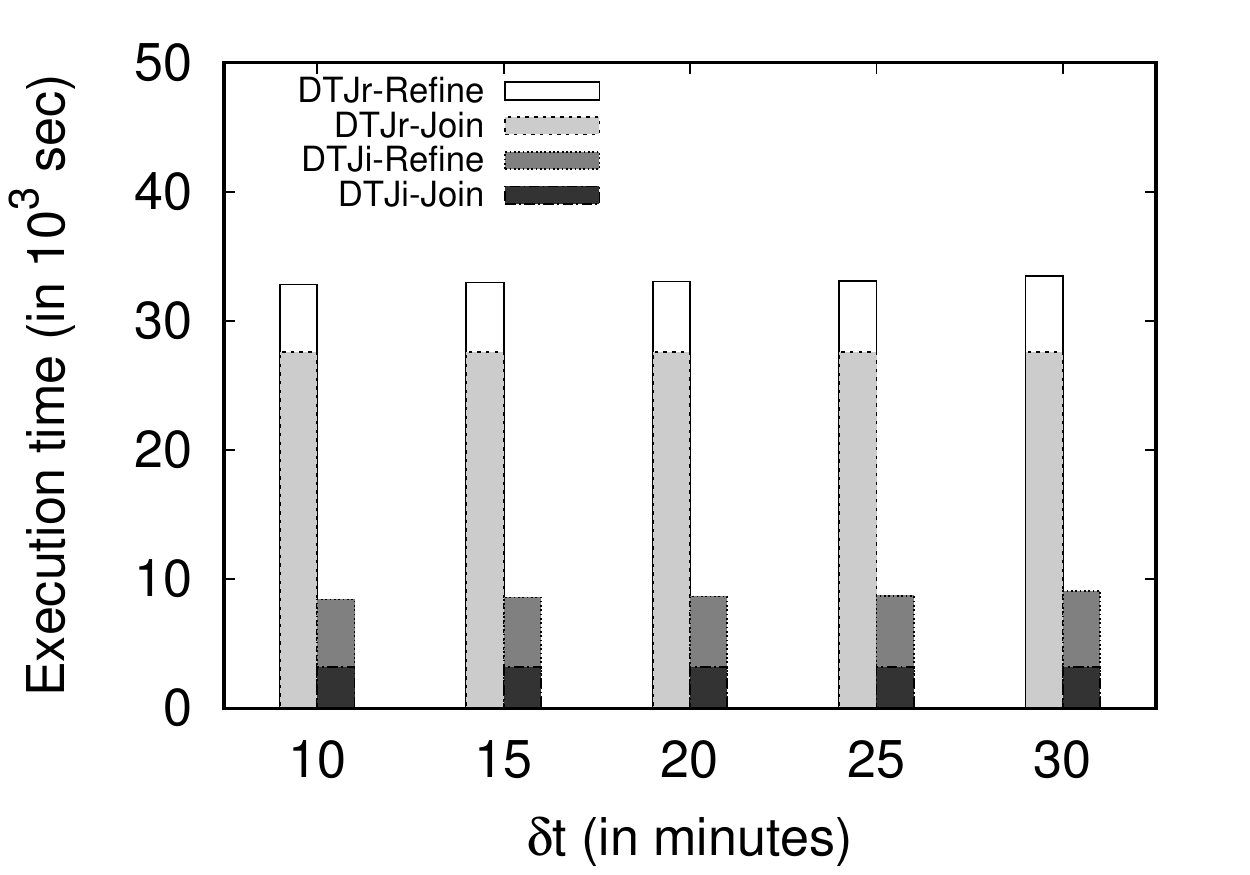}
  \end{minipage}
  \caption{Sensitivity analysis varying (a) $\epsilon_t$, (b) $\epsilon_{sp}$ and (c) $\delta{t}$}
  \label{fig_Sensitivity}
\end{figure*}

Initially we vary the value of $\epsilon_t$ while retaining fixed the values of the other parameters. As shown in Figure~\ref{fig_Sensitivity}(a), the execution time of both algorithms, as expected, increases with $\epsilon_t$. In more detail, the  \emph{Join} phase of $\rep$ is more sensitive to the fluctuation $\epsilon_t$ than the  \emph{Join} phase of $\ind$, due to the fact that the latter is utilizing the SpI index which, for a given $\epsilon_t$, performs spatial filtering instead of scanning the entire space in order to find ``matching'' pairs of points. What is more interesting is that as $\epsilon_t$ increases the difference between the two approaches increases, which means that for higher values of $\epsilon_t$ the difference, in terms of execution time, will be higher than one order of magnitude. As far as it concerns the  \emph{Refine} step, both approaches present the same increasing behavior when $\epsilon_t$ increases, since both of them employ the same algorithm, due to the fact that the higher the $\epsilon_t$, the larger the sliding window that is created.

Then, we set different values to $\epsilon_{sp}$ while keeping the values of the other parameters fixed. As illustrated in Figure~\ref{fig_Sensitivity}(b), $\epsilon_{sp}$ affects directly the  \emph{Join} and indirectly the  \emph{Refine} phase of both approaches. More specifically, the  \emph{Join} phase of $\rep$ is slightly affected by setting different values to $\epsilon_{sp}$ due to the fact that, for a given $\epsilon_t$, $\rep$ will search the whole space in order to find ``matches''. Hence, $\epsilon_{sp}$ will only affect the number of ``matches''. On the other hand, $\ind$ does not search the whole space but utilizes the SpI index which, consequently, makes it more sensitive to $\epsilon_{sp}$. The only case where the  \emph{Join} phase of $\rep$ performs the same as $\ind$ is when $\epsilon_{sp}$ spans the whole dataset space. Regarding the \emph{Refine} step, as expected, both approaches perform the same and the higher the $\epsilon_{sp}$, the higher the execution time. The reason for this behaviour is that as $\epsilon_{sp}$ increases, the product of the  \emph{Join} phase increases. 

Finally, we vary the values of $\delta{t}$ while keeping the values of the other parameters fixed. As presented in Figure~\ref{fig_Sensitivity}(c), this parameter affects only the \emph{Refine} phase, as anticipated. More specifically, the higher the $\delta{t}$ the slightly higher the execution time of both approaches. This takes place due to the fact that as $\delta{t}$ increases, the sliding window gets larger. 

\subsection{Indexing} \label{sec_idx}

\begin{figure*} [thb]
  \begin{minipage}[t]{1\linewidth}
  \centering
  (a)\includegraphics[width=.45\linewidth]{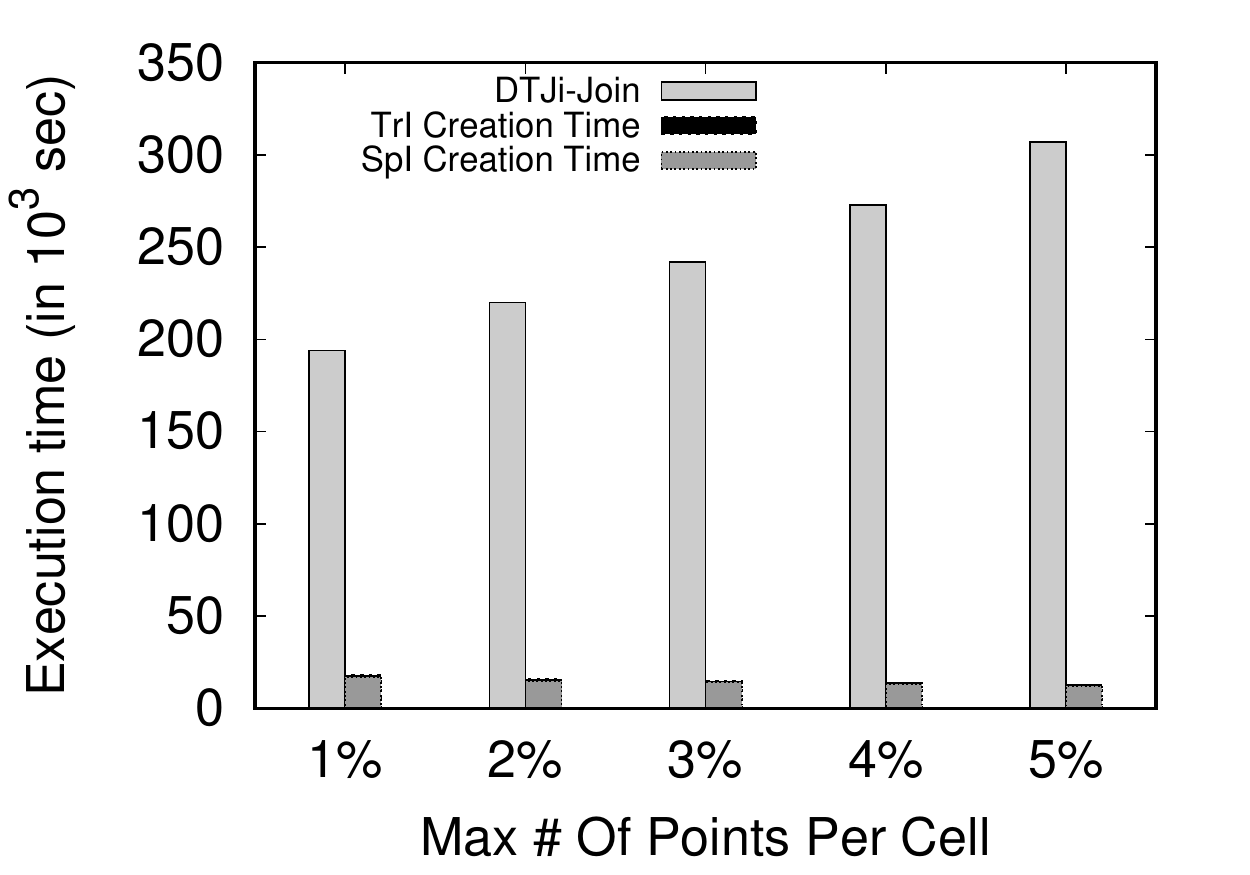}
  (b)\includegraphics[width=.45\linewidth]{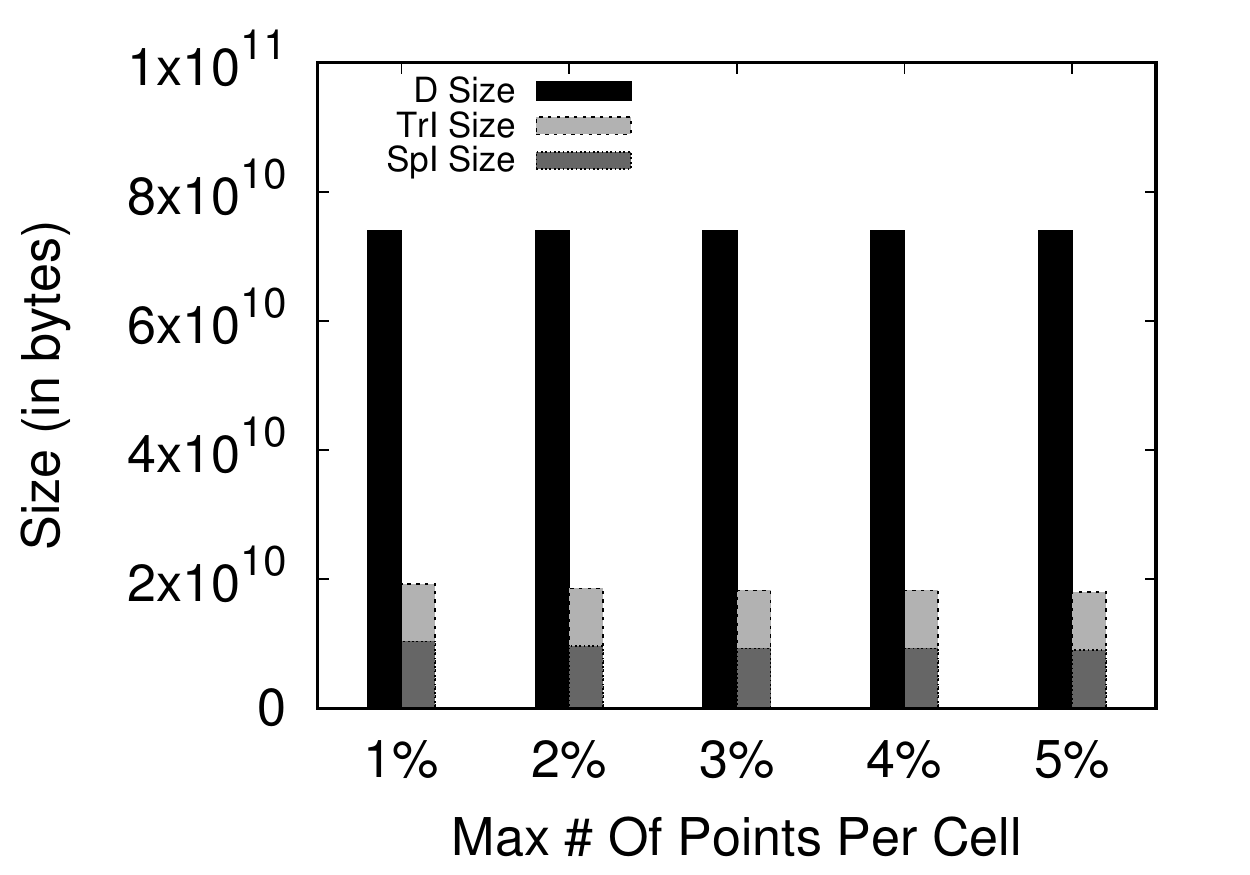}
  (c)\includegraphics[width=.45\linewidth]{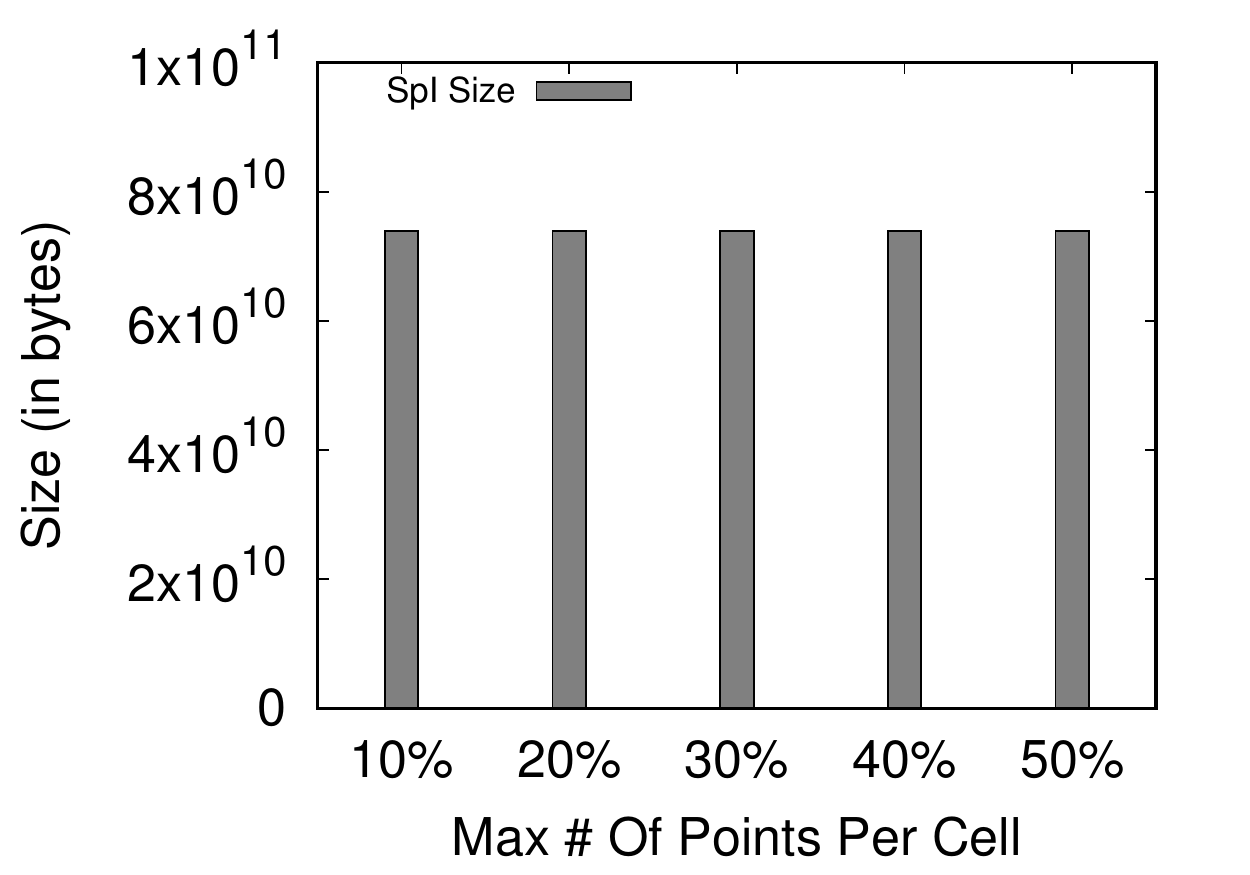}
  \end{minipage}
  \caption{(a) Index construction time, (b) Index size and (c) Effect of $\epsilon_{sp}$}
  \label{fig_Indexing}
\end{figure*}

In order to measure the effect of having different number of spatial partitions in spatial index size and spatial index construction time, we perform a final set of experiments. More specifically, we vary the maximum number of points per cell parameter of the QuadTree, and we measure the index creation time and the index size. As illustrated in Figure~\ref{fig_Indexing}(a), the SpI construction time increases as the maximum number of points per cell decrease, while the TrI construction time is, as expected, not affected by that. This occurs due to the fact that as the maximum number of points per cell decreases, the number of spatial partitions increases. Furthermore, as depicted in Figure~\ref{fig_Indexing}(a), the fewer the maximum number of points per cell the smaller the execution time of the  \emph{Join} algorithm. It is worth mentioning that the overall index construction time as a percentage over the execution time of the  \emph{Join} algorithm varies only between 4\% and 11\%.

As far as the size of the indexes is concerned, Figure~\ref{fig_Indexing}(b) illustrates how the size is affected when varying the maximum number of points per cell. As expected, the TrI index is not affected, whereas the SpI index slightly increases its size as the number of partitions increase. At this point, we should mention that compared to the size of $D$, the percentage of the total size of the indexing scheme over $D$ varies only between 24\% and 28\%.

Another parameter that can affect the SpI index is $\epsilon_{sp}$ due to the fact that each spatial partition is enlarged by $\epsilon_{sp}$. As depicted in Figure~\ref{fig_Indexing}(c), the size of SpI increases as $\epsilon_{sp}$ increases.

\section{Related Work} \label{sec_relat}

The work performed in this paper is closely related to three topics in the spatial and spatio-temporal database management literature. These are, (a) centralized trajectory joins, (b) MapReduce spatial and multidimensional joins, and (c) distributed trajectory joins. In the following paragraphs we outline the related work and try to highlight their differences from our work. 

\textbf{Centralized Trajectory Joins:} In~\cite{DBLP_conf/mdm/BakalovHKT05} the effort focuses in identifying pairs of trajectories that move close enough, with respect to a spatial threshold, during a user specified temporal window, unlike our approach where we identify ``matches'' of different duration (at least $\delta t$) during the whole lifespan of the datasets. Furthermore, in~\cite{DBLP_conf/mdm/BakalovHKT05}, no temporal tolerance is considered which can lead in missing pairs of trajectories that move with some temporal displacement. In ~\cite{DBLP_conf/gsn/BakalovT06}, the authors try to solve the same problem in a streaming environment. In~\cite{DBLP_conf/gis/BakalovHT05} the authors extend their work by not binding to the temporal dimension the interval in which two trajectories should move ``together''. Hence, all pairs of ``matching'' (w.r.t a spatial threshold) subtrajectories of exactly $\delta t$ duration will be returned. This definition, although more general, it still suffers from the rest of the aforementioned problems. Moreover, in these approaches there is an assumption made, that all trajectories have the same number of points which are synchronized. However, an assumption like that is not realistic in real life scenarios and supposes a preprocessing step that can be prohibitive when dealing with Big Data.  A slightly different definition is provided in~\cite{DBLP_conf/gis/ChenP09} where the goal is to identify all pairs of moving objects that, for some time intervals, move closer than a given spatial threshold. Here, the duration of the ``matches'' is not fixed. The most significant difference with our approach is that, here, they identify only the \emph{joining points}, which leads in a trajectory join definition which is asymmetric. Moreover, time relaxation is not considered, so the distance between two objects refers to their distance at the same time point $t$. Furthermore, the minimum duration of the ``matches'' cannot be limited which can lead to pairs with very small duration that might not be useful for some applications. Finally, the solution provided is focused to an instantiation of the problem, called Window Trajectory Distance Join, which limits the problem to a user-specified time window. 

A similar but different problem is the one of trajectory similarity join, where the goal is to retrieve all pairs of trajectories that exceed a given similarity threshold as in ~\cite{DBLP_journals/tkde/TaLXLHF17} and ~\cite{DBLP_conf/time/DingTS08}. However, both of them return as a result pairs of trajectories and not subtrajectories, thus they cannot support some of the scenarios presented in Section~\ref{sec_intro}. An approach very similar to ours is presented in ~\cite{DBLP_journals/comgeo/BuchinBKL11}, where, given a pair of trajectories they try to perform partial matching, finding the most similar subtrajectories between these two
trajectories. Different variations of the problem are presented, where the duration of the ``match'' is specified beforehand or not. Nevertheless, the problem in~\cite{DBLP_journals/comgeo/BuchinBKL11} is not a join operation and temporal tolerance is not considered. To sum up, all of the above approaches are centralized and applying them to a parallel and distributed environment is non-trivial. 

\textbf{MapReduce Spatial \& Multidimensional Joins:} A special class of joins which is very relative to our problem is that of spatial join. There have been several efforts to tackle this issue using the MapReduce framework. In particular, \cite{DBLP_conf/cluster/ZhangHLWX09}, which is based on the traditional PPBSM algorithm~\cite{DBLP_conf/sigmod/PatelD96}, partitions the input data into small evenly disjoint tiles at \emph{Map} stage and joins them at \emph{Reduce} stage by further partitioning the data into strips and performing a plane sweeping algorithm along with a duplication avoidance technique.
Other works that try to deal with the problem of spatial join are ~\cite{DBLP_journals/pvldb/AjiWVLL0S13} and~\cite{DBLP_conf/icde/EldawyM15}. In more detail, ~\cite{DBLP_journals/pvldb/AjiWVLL0S13} first partitions the data by focusing on recursively breaking high density tiles into smaller ones. Objects that exceed the borders of a partition are replicated and a post-processing step is employed in order to eliminate duplicate results. The join process takes place at the \emph{Reduce} phase by utilizing the R*-Tree indexes that are created and loaded in-memory at query time for each of the relations. Another system that copes with the problem of spatial join is~\cite{DBLP_conf/icde/EldawyM15}. This approach first re-partitions the data by taking into account load balancing and spatial locality. In more detail the data are sampled and an index (Grid File, R-Tree or R+-Tree) is created which will set the boundaries of each partition. During, the re-partitioning phase global and local indexes are created and stored in HDFS.  The join takes place in the \emph{Map} phase by utilizing the local indexes. Concerning borderline objects, a duplicate avoidance method is applied. 
An approach that enhances~\cite{DBLP_conf/icde/EldawyM15} with the functionality of identifying closest pairs of points is presented in~\cite{DBLP_conf/adbis/Garcia-GarciaCI16}.
Multidimensional similarity join is also related to our work. In ~\cite{DBLP_conf/cloudi/SilvaRT12,DBLP_conf/sigmod/SilvaR12} the problem of distance range join is studied, which is probably the most common case of similarity join. In this approach the data is iteratively partitioned similarly to the Quickjoin algorithm~\cite{DBLP_journals/tods/JacoxS08}, which results in having multiple MR jobs in order to get the final results. 
The problem of high dimensional similarity joins on massive datasets using MapReduce is tackled in~\cite{DBLP_conf/mdm/LuoTMN12}.
In~\cite{DBLP_conf/btw/SeidlFB13} the problem of $\epsilon$-distance similarity self-join on vector data is tackled by employing in the \emph{Map} phase a fixed size grid with cell width $\epsilon$ and assign the data to the corresponding cells. In order to compute the $\epsilon$-neighborhood of each cell only the adjacent cells are needed. In order to reduce the replication of data they avoid taking into account all the adjacent cells. Despite this, the algorithm used to perform the join in the \emph{Reduce} phase is still a Nested Loop Join. \cite{DBLP_conf/icde/FriesBSS14} is an extension of~\cite{DBLP_conf/btw/SeidlFB13} for medium- to high-dimensional spaces where the full $d$-dimensional space is broken down to $k$ dimension groups, the join is performed in each group and then the results are merged. Unlike~\cite{DBLP_conf/btw/SeidlFB13}, they try to cope with the skewness of data by starting with a very fine grid and merging cells until a balanced grid is created. Similarly to~\cite{DBLP_conf/btw/SeidlFB13} the join process is still a Nested Loop Join.
However, all of the above approaches try to solve a problem that is significantly different from ours since our problem is not to join spatial or multidimensional objects but to join sequences of spatial points. In more detail, the output of the above approaches would be the set of $JP$. However, as already stated, in order to get the desired result we also need to the sets of $BP$ and $SNJP$, which cannot be acquired by post-processing the result of these approaches.

\textbf{Distributed Trajectory Joins:} The approaches mainly focus to the $k$-nn join or the $k$ most similar trajectories join. More specifically, ~\cite{DBLP_conf/cikm/Zeinalipour-YaztiLG06} and ~\cite{DBLP_journals/pvldb/XieLP17} address the problem where given a reference trajectory and an integer $k$ they want to discover the $k$ most similar trajectories to the reference trajectory. 
The problem of $k$-nn join by using the MapReduce framework is tackled in ~\cite{DBLP_conf/icde/FangCTMY16}. More specifically, given two sets of trajectories $R$ and $M$, an integer $k$ and a time interval $[t_s, t_e]$, the algorithms proposed there return the k nearest neighbors from $R$ for each object in $M$ during this interval. The above approaches, address an entirely different problem than the one presented in this paper.
Recently, the algorithms proposed in~\cite{DBLP_journals/pvldb/ShangCWJ0K17,DBLP_journals/vldb/ShangCWJZK18} find all pairs of network-constrained trajectories that exceed a similarity threshold in a parallel manner. However, the parallelization proposed there handles each trajectory separately by assuming that all data need to be replicated for each trajectory, which makes such a solution inapplicable to the Big Data setting. Finally, these approaches (a) assume that the underlying network is known, which is not something trivial in some domains (e.g. maritime or aviation) and (b) work at the entire trajectories and cannot identify matching sub-trajectories. 

\section{Conclusions} \label{sec_concl}

In this paper, we introduced the \emph{\prob} query, an important operation in the spatiotemporal data management domain, where very large datasets of moving object trajectories are processed for analytic purposes. To address this problem in an efficient manner following the MapReduce programming model, we initially provided a well-designed basic solution which is used as a baseline in order to propose two efficient improvements, called $\rep$ and $\ind$ which can boost the performance by up to 16$\times$ and 10$\times$, respectively. Our experimental study was performed on a very large real dataset of trajectories from the maritime domain,consisting of 56 GB of data (or 1.5 billion time-stamped locations). As for future work, the solution to the \emph{\prob} problem that we provided can be utilized as the basis for efficiently identifying various mobility patterns (e.g. group behaviours) and discovering clusters of moving objects over massively distributed data. 

\section{Acknowledgements} \label{sec_ack}

This work was partially supported by projects datACRON (grant agreement No 687591), Track\&Know (grant agreement No 780754) and MASTER (Marie Sklowdoska-Curie agreement N. 777695), which have received funding from the EU Horizon 2020 R\&I Programme. We would like to thank the Greek Research and Technology Network (GRNET) for providing access to the Okeanos cloud infrastructure and IMIS Hellas for donating the IMIS dataset.

\bibliographystyle{abbrv}

\bibliography{main}
\appendix

\section{Appendix}
\subsection*{Proof of Lemma~\ref{lem_breaking}}
\begin{proof}
It suffices to construct an instance of the problem where an algorithm $A$ that operates solely on \emph{joining points} and is unaware of \emph{breaking points} would produce erroneous results, thus $A \notin \mathcal{A}$.
Let $r'= r_{1,n}$ and $s'= s_{1,(n+1)}$ denote two subtrajectories, such that there exist $n$ pairs of \emph{joining points} ($r_i$,$s_j$) and $\Delta w_{r',s'} = \delta t$. However, let us assume that there exists a point $s_k \in s'$ which is a \emph{breaking point}. Based on the problem definition (Defn.~\ref{dfn_3}), if algorithm $A$ was unaware of \emph{breaking points}, it would falsely identify $r'$ and $s'$ as ``matching'' subtrajectories.
\end{proof}

\subsection*{Proof of Lemma~\ref{lem_nonjoining}}
\begin{proof}
The proof is similar to the proof of Lemma~\ref{lem_breaking}, only using a \emph{non-joining point} instead of a \emph{breaking point} in the constructed instance of the problem.
\end{proof}

\subsection*{Proof of Lemma~\ref{lem_partitioning1}}
\begin{proof}
\textbf{Joining points}: By contradiction. Let us assume that an expanded partition $expPart_i$ is not sufficient to produce the set of $JP$ in $part_i$. Then, there must exist a pair of joining points $r_j \in r$ and $s_k \in s$, such that $r_j$ belongs to partition $part_i$, whereas $s_k$ does not belong to $expPart_i$. Based on the definition of expanded partitions, it follows that  $DistT(r_j, s_k) > \epsilon_t$. Thus, $r_j$ and $s_k$ cannot be joining points, which is a contradiction.
\textbf{Breaking points}: By contradiction. Let us assume that an expanded partition $expPart_i$ is not sufficient to produce the set of $BP$ for trajectories in $part_i$. Then, there must exist a point $r_*$ (that belongs to $part_i$) of trajectory $r$, such that $r_*$ is a breaking point. To identify if $r_*$ is a breaking point, we need to examine whether there exists a point $s_j \in s$ of any other trajectory $s$ with $DistT(r_*, s_j) \leq \epsilon_t$ (Defn.~\ref{dfn_6}). However, based on the definition of expanded partitions, such a point $s_j$ must belong to $expPart_i$, which contradicts with the assumption that $expPart_i$ is not sufficient.
\end{proof}

\subsection*{Proof of Lemma~\ref{lem_partitioning2}}
\begin{proof}
Consider two pairs of $JP$, $(r_{i},s_{j})$ and $(r_{i+1},s_{j+n})$, where $r_{i},r_{i+1} \in r$ and $s_{j},s_{j+n} \in s$. Moreover, let us assume that for each point $s_{m}$ where $m \in [j+1, j+n-1]$, $s_{m}$ is a \emph{non-joining point} for each point of $r$. Then, points $r_{i}$, $r_{i+1}$, $s_{j}$, $s_{m}$ and $s_{j+n}$ will span at most to two consecutive temporal partitions: $part_{k}$ and $part_{k+1}$. This means that either $s_{j}, s_{m} \in part_{k}$ or $s_{m}, s_{j+n} \in part_{k+1}$. In both cases $s_{m}$ will be recognized as a \emph{non-joining point} for each point of $r$, by the procedure that generates the set of $sNJP$(Defn.~\ref{dfn_8}).  
\end{proof}

\end{document}